\documentclass[twocolumn]{aastex63}
\received{Month 00, 2020}
\revised{Month 00, 2021}
\accepted{Month 00, 2021}
\submitjournal{ApJL}

\shorttitle{TiO in hot Jupiter HAT-P-65~b}
\shortauthors{Chen et al.}

\begin{document}

\title{Evidence for TiO in the atmosphere of the hot Jupiter HAT-P-65~b}

\correspondingauthor{Guo Chen}
\email{guochen@pmo.ac.cn}

\author[0000-0003-0740-5433]{Guo Chen}
\affiliation{CAS Key Laboratory of Planetary Sciences, Purple Mountain Observatory, Chinese Academy of Sciences, Nanjing 210023, People's Republic of China}
\affiliation{CAS Center for Excellence in Comparative Planetology, Hefei 230026, People's Republic of China}

\author[0000-0003-0987-1593]{Enric Pall\'{e}}
\affiliation{Instituto de Astrof\'{i}sica de Canarias, V\'{i}a L\'{a}ctea s/n, E-38205 La Laguna, Tenerife, Spain}
\affiliation{Departamento de Astrof\'{i}sica, Universidad de La Laguna, E-38206 La Laguna, Tenerife, Spain}

\author[0000-0001-5519-1391]{Hannu Parviainen}
\affiliation{Instituto de Astrof\'{i}sica de Canarias, V\'{i}a L\'{a}ctea s/n, E-38205 La Laguna, Tenerife, Spain}
\affiliation{Departamento de Astrof\'{i}sica, Universidad de La Laguna, E-38206 La Laguna, Tenerife, Spain}

\author[0000-0001-9087-1245]{Felipe Murgas}
\affiliation{Instituto de Astrof\'{i}sica de Canarias, V\'{i}a L\'{a}ctea s/n, E-38205 La Laguna, Tenerife, Spain}
\affiliation{Departamento de Astrof\'{i}sica, Universidad de La Laguna, E-38206 La Laguna, Tenerife, Spain}

\author[0000-0001-9585-9034]{Fei Yan}
\affiliation{Institut f\"{u}r Astrophysik, Georg-August-Universit\"{a}t, Friedrich-Hund-Platz 1, D-37077 G\"{o}ttingen, Germany}



\begin{abstract}

We present the low-resolution transmission spectra of the puffy hot Jupiter HAT-P-65b (0.53~M$_\mathrm{Jup}$, 1.89~R$_\mathrm{Jup}$, $T_\mathrm{eq}=1930$~K), based on two transits observed using the OSIRIS spectrograph on the 10.4~m Gran Telescopio CANARIAS (GTC). The transmission spectra of the two nights are consistent, covering the wavelength range 517--938~nm and consisting of mostly 5~nm spectral bins. We perform equilibrium-chemistry spectral retrieval analyses on the jointly fitted transmission spectrum and obtain an equilibrium temperature of $1645^{+255}_{-244}$~K and a cloud coverage of $36^{+23}_{-17}$\%, revealing a relatively clear planetary atmosphere. Based on free-chemistry retrieval, we report strong evidence for TiO. Additional individual analyses in each night reveal weak-to-moderate evidence for TiO in both nights, but moderate evidence for Na or VO only in one of the nights. Future high-resolution Doppler spectroscopy as well as emission observations will help confirm the presence of TiO and constrain its role in shaping the vertical thermal structure of HAT-P-65b's atmosphere. 

\end{abstract}

\keywords{Exoplanet atmospheres --- Exoplanet atmospheric composition --- Transmission spectroscopy --- Hot Jupiters --- Exoplanets}


\section{Introduction}
\label{sec:intro}

Titanium and vanadium oxides (TiO and VO) exhibit prominent molecular absorption bands in the optical spectra of M dwarfs, and their signatures gradually disappear in cooler L dwarfs \citep[$T_\mathrm{eff}\sim$ 1700--1900~K;][]{1999ApJ...519..802K}. Similarly, TiO and VO have long been expected to be the dominant opacity sources in the atmospheres of highly irradiated hot Jupiters and responsible for causing thermal inversions \citep{2003ApJ...594.1011H,2008ApJ...678.1419F}. 

Up until now, thermal inversions have been detected in the dayside of several ultra-hot Jupiters using low-resolution emission observations \citep[e.g.,][]{2015ApJ...806..146H,2017Natur.548...58E,2017ApJ...850L..32S,2018ApJ...855L..30A,2018AJ....156...10M}. Recent high-resolution Doppler emission spectroscopy observations \citep{2020ApJ...894L..27P,2020A&A...640L...5Y,2020ApJ...898L..31N} reveal that these inversions are at least partly induced by optical absorbers such as atomic Fe or Mg lines or continuum H$^{-}$, regardless of the presence of TiO or VO \citep{2018ApJ...866...27L}. 

Indeed, TiO and VO have been rarely detected. The most confident detection of TiO comes from the atmosphere of WASP-33b with high-resolution Doppler emission spectroscopy \citep[][but also see \citealt{2020AJ....160...93H} and \citealt{2020arXiv201110587S}]{2017AJ....154..221N}. \citet{2017Natur.549..238S} reported a detection of TiO in the low-resolution transmission spectrum of WASP-19b, while \citet{2019MNRAS.482.2065E} reported a non-detection based on independent observations. Tentative evidences of TiO/VO were also claimed in the low-resolution transmission spectrum of WASP-121b \citep{2016ApJ...822L...4E,2018AJ....156..283E}, but later \citet{2020A&A...636A.117M} reported non-detections for both TiO and VO using high-resolution Doppler transmission spectroscopy. Thus, the existence and detectability of TiO/VO in hot-Jupiter atmospheres remain an open question.

Here, we report strong evidence for TiO in the atmosphere of the hot Jupiter HAT-P-65b using low-resolution transmission spectroscopy. HAT-P-65b is a very puffy hot Jupiter (0.53~M$_\mathrm{Jup}$, 1.89~R$_\mathrm{Jup}$) orbiting a G2 star every 2.61~days \citep{2016AJ....152..182H}. With an equilibrium temperature of $1930\pm 45$~K, it falls near the transition regime where ultra-hot Jupiters are defined \citep[e.g., $T_\mathrm{eq}>2000$~K;][]{2018ApJ...866...27L}. In Section \ref{sec:obs}, we summarize the observations and data reduction. We then describe the light-curve analysis in Section \ref{sec:lc} and interpret the transmission spectrum in Section \ref{sec:transpec}. Finally, we give the summary and discuss its implications in Section \ref{sec:discuss}.

\section{Observations and data reduction}
\label{sec:obs}

We observed two transits of \object{HAT-P-65b} under the programs GTCMULTIPLE2C-18A (PI: G. Chen) and GTC24-20A (PI: E. Pall\'{e}) using the OSIRIS spectrograph \citep{2000SPIE.4008..623C} installed on the 10.4~m Gran Telescopio CANARIAS (GTC) in La Palma, Spain. OSIRIS is equipped with two CCDs to cover the unvignetted field of view of $\sim$7.5$'$. In both observations, OSIRIS was configured in the long-slit mode with a 12$''$ slit. The target star \object{HAT-P-65} ($r'=12.95$~mag) and the reference star \object{2MASS J21034527+1158213} ($r'=12.88$~mag, 2.19$'$ away) were both located on CCD 1 and simultaneously placed in the slit. The 200~kHz readout and 2$\times$2 binning (0.254$''$ per binned pixel) were adopted. The R1000R grism was used to acquire the spectra, covering a wavelength range of 510--1000~nm at a dispersion of $\sim$2.6~\AA~per binned pixel. The HeAr, Ne, and Xr arc lamps were observed through a 1.23$''$ slit.

The two transits were observed on the nights of 2018 July 29 (Night 1) and 2020 August 7 (Night 2), covering a UT window of 21:27--05:41 and 21:08--04:05, respectively. The typical seeing was $\sim$0.9$''$ for Night 1 and $\sim$0.6$''$ for Night 2. To avoid saturation, a slight defocusing was used in both nights, increasing the full widths at half maximum (FWHMs) of the point spread function (PSF) in the cross-dispersion direction to a median value of $\sim$1.3$''$ (Night 1) and $\sim$1.0$''$ (Night 2). An exposure time of 75 sec was used for both nights, resulting in 305 and 255 spectra for the first and second nights, respectively. For Night 1, there was a thin cirrus crossing event after the ingress ended. Nine spectra acquired during that time were discarded. 

We reduced the raw data using the methodology adopted in our previous studies \citep[e.g.,][]{2018A&A...616A.145C,2020A&A...642A..54C}. The spectral images were corrected for overscan, bias, flat, sky background, and cosmic rays. The one dimensional spectra were extracted using an aperture radius of 9 pixels (Night 1) and 8 pixels (Night 2). The aperture radius was optimized by minimizing the point-to-point scatter in the white light curves. We note that there is a faint background companion star at $\sim$3.6$''$ but contributing negligible flux within the chosen apertures (see Appendix \ref{sec:dilution}). The initial wavelength solution was constructed by the arc lines, which was later revised to allow that all the spectra of the target and reference stars were well aligned in the wavelength domain. The time stamp was created in Barycentric Julian Dates in Barycentric Dynamical Time \citep[$\mathrm{BJD}_\mathrm{TDB}$;][]{2010PASP..122..935E}.

We created the white light curves by summing the flux within the wavelength range 514--908~nm, but excluding the 754--768~nm region to avoid the telluric oxygen A-band. To create the spectroscopic light curves, we divided the spectra into 63 channels of 5~nm bin, six channels of 11~nm bin, and one channel of 31~nm bin. We set up broader bins at longer wavelengths due to the appearance of fringing pattern at $\lambda\gtrsim 830$~nm.

\section{Light-curve analysis}
\label{sec:lc}

We modeled the white and spectroscopic light curves following the approach described in \citet{2018A&A...616A.145C,2020A&A...642A..54C}. Here we first give a summary of general procedures, and then present the specifics in the subsections. 

The light curves were assumed to consist of astrophysical signals and correlated systematics. The transit was described by the \citet{2002ApJ...580L.171M} model, implemented via the Python package \texttt{batman} \citep{2015PASP..127.1161K} and parameterized by orbital period $P$ \citep[fixed to 2.6054552~d;][]{2016AJ....152..182H}, orbital inclination $i$, scaled semi-major axis $a/R_\star$, radius ratio $R_\mathrm{p}/R_\star$, mid-transit time $T_\mathrm{mid}$, and limb-darkening coefficients $u_i$. A circular orbit was adopted \citep{2016AJ....152..182H}. The correlated systematics were treated as Gaussian processes \citep[GP;][]{2006gpml.book.....R,2012MNRAS.419.2683G}, implemented via the Python package \texttt{george} \citep{2015ITPAM..38..252A}. 

The quadratic limb-darkening law was adopted, and the coefficients $u_1$ and $u_2$ were fitted with Gaussian priors. The priors were calculated using the Kurucz ATLAS9 stellar atmosphere models by a Python code from \citet{2015MNRAS.450.1879E}. The grid with stellar effective temperature $T_\mathrm{eff}=5750$~K, surface gravity $\log g_\star=4.0$, and metallicity $\mathrm{[Fe/H]}=0.1$ was used to derive the prior mean values, while the average differences among the grids with $T_\mathrm{eff}=5500$~K, 5750~K, and 6000~K were adopted as the prior sigma values. 

To explore the marginalized posterior distributions of fitted parameters, the affine-invariant Markov chain Monte Carlo (MCMC) was used, implemented via the Python package \texttt{emcee} \citep{2013PASP..125..306F}. The walkers are initialized around literature values for transit parameters and arbitrary values for others. The number of walkers are set to be at least twice the number of free parameters (50 and 32 for the white and spectroscopic light curves, respectively). We always run two short chains (1000 steps) for the 'burn-in' phase and then start a long chain (20000 steps) for the posterior estimation. The chain length is kept to be more than 50 times autocorrelation time to ensure convergence.

\begin{deluxetable}{lcr}
\tablecaption{Parameters derived from white light curves. \label{tab:params}}
\tablewidth{0pt}
\tabletypesize{\footnotesize}
\tablehead{
\colhead{Parameter} & 
\colhead{Prior} & 
\colhead{Posterior estimate}
} 
\startdata
    $P$ [d]                             & 2.6054552(fixed)               & -- \\ \noalign{\smallskip}
    $i$ [$^{\circ}$]                    & $\mathcal{U}(80,90)$           &  $89.10^{+0.63}_{-0.83}$ \\ \noalign{\smallskip}
    $a/R_\star$                         & $\mathcal{U}(2,8)$             &  $5.221^{+0.025}_{-0.043}$ \\ \noalign{\smallskip}
    $R_\mathrm{p}/R_\star$              & $\mathcal{U}(0.05,0.15)$       &  $0.0994^{+0.0025}_{-0.0025}$ \\ \noalign{\smallskip}
    $u_1$                               & $\mathcal{N}(0.356,0.045^2)$   &  $0.333^{+0.037}_{-0.038}$ \\ \noalign{\smallskip}
    $u_2$                               & $\mathcal{N}(0.277,0.027^2)$   &  $0.270^{+0.026}_{-0.025}$ \\ \noalign{\smallskip}
    \hline\noalign{\smallskip}
    \multicolumn{3}{c}{2018-07-29: Night 1} \\ \noalign{\smallskip}
    $T_\mathrm{mid}$ [MJD$\tablenotemark{a}$] & $\mathcal{U}(8329.53,8329.57)$  &  $8329.54828^{+0.00029}_{-0.00029}$ \\ \noalign{\smallskip}
    $\sigma_w$ [$10^{-6}$]              & $\mathcal{U}(0.1,5000)$        &  $284^{+16}_{-15}$ \\ \noalign{\smallskip}
    $\ln A$                             & $\mathcal{U}(-10,-1)$          &  $-6.41^{+0.36}_{-0.28}$ \\ \noalign{\smallskip}
    $\ln\tau_t$                         & $\mathcal{U}(-6,6)$            &  $-2.52^{+0.33}_{-0.29}$ \\ \noalign{\smallskip}
    $\ln\tau_x$                         & $\mathcal{U}(-6,6)$            &  $1.64^{+0.41}_{-0.35}$ \\ \noalign{\smallskip}
    \hline\noalign{\smallskip}
    \multicolumn{3}{c}{2020-08-07: Night 2} \\ \noalign{\smallskip}
    $T_\mathrm{mid}$ [MJD$\tablenotemark{a}$] & $\mathcal{U}(9069.48,9069.52)$ &  $9069.49472^{+0.00034}_{-0.00034}$ \\ \noalign{\smallskip}
    $\sigma_w$ [$10^{-6}$]              & $\mathcal{U}(0.1,5000)$        &  $147^{+12}_{-12}$ \\ \noalign{\smallskip}
    $\ln A$                             & $\mathcal{U}(-10,-1)$          &  $-6.74^{+0.33}_{-0.24}$ \\ \noalign{\smallskip}
    $\ln\tau_t$                         & $\mathcal{U}(-6,6)$            &  $-3.32^{+0.27}_{-0.23}$ \\ \noalign{\smallskip}
    $\ln\tau_x$                         & $\mathcal{U}(-6,6)$            &  $3.52^{+1.39}_{-0.84}$ \\ \noalign{\smallskip}
\enddata
\tablenotetext{a}{$\mathrm{MJD}=\mathrm{BJD}_\mathrm{TDB}-2450000$.}
\end{deluxetable}

\subsection{White light curves}

The white light curves of the two nights were jointly fitted, sharing the same values for ($i$, $a/R_\star$, $R_\mathrm{p}/R_\star$, $u_1$, $u_2$). The transit model was adopted as the GP mean function, while the Mat\'{e}rn $\nu=3/2$ kernel
\begin{equation}
k_{ij}= A^2\left( 1+\sqrt{3\,D_{ij}^2} \right)\,\exp \left (-\sqrt{3\,D_{ij}^2} \right)
\end{equation}
was adopted as the GP covariance matrix, where $D_{ij}^2=\sum_{\alpha=1}^{N}{(\hat{x}_{\alpha,i}-\hat{x}_{\alpha,j})^2/\tau_\alpha^2}$, and $A$ and $\tau_\alpha$ are the characteristic correlation amplitude and length scale. We chose time $t$ and spectral drift $x$ as the GP covariance matrix inputs (see Appendix \ref{sec:modelselect} for model selection details). In the light-curve modeling, uniform priors were adopted for ($i$, $a/R_\star$, $R_\mathrm{p}/R_\star$, $T_\mathrm{mid}$, $\sigma_w$), while log-uniform priors were adopted for ($A$, $\tau_t$, $\tau_x$), where $\sigma_w$ is the white noise jitter to inflate the light-curve uncertainties.

\begin{figure*}
\centering
\includegraphics[width=\textwidth]{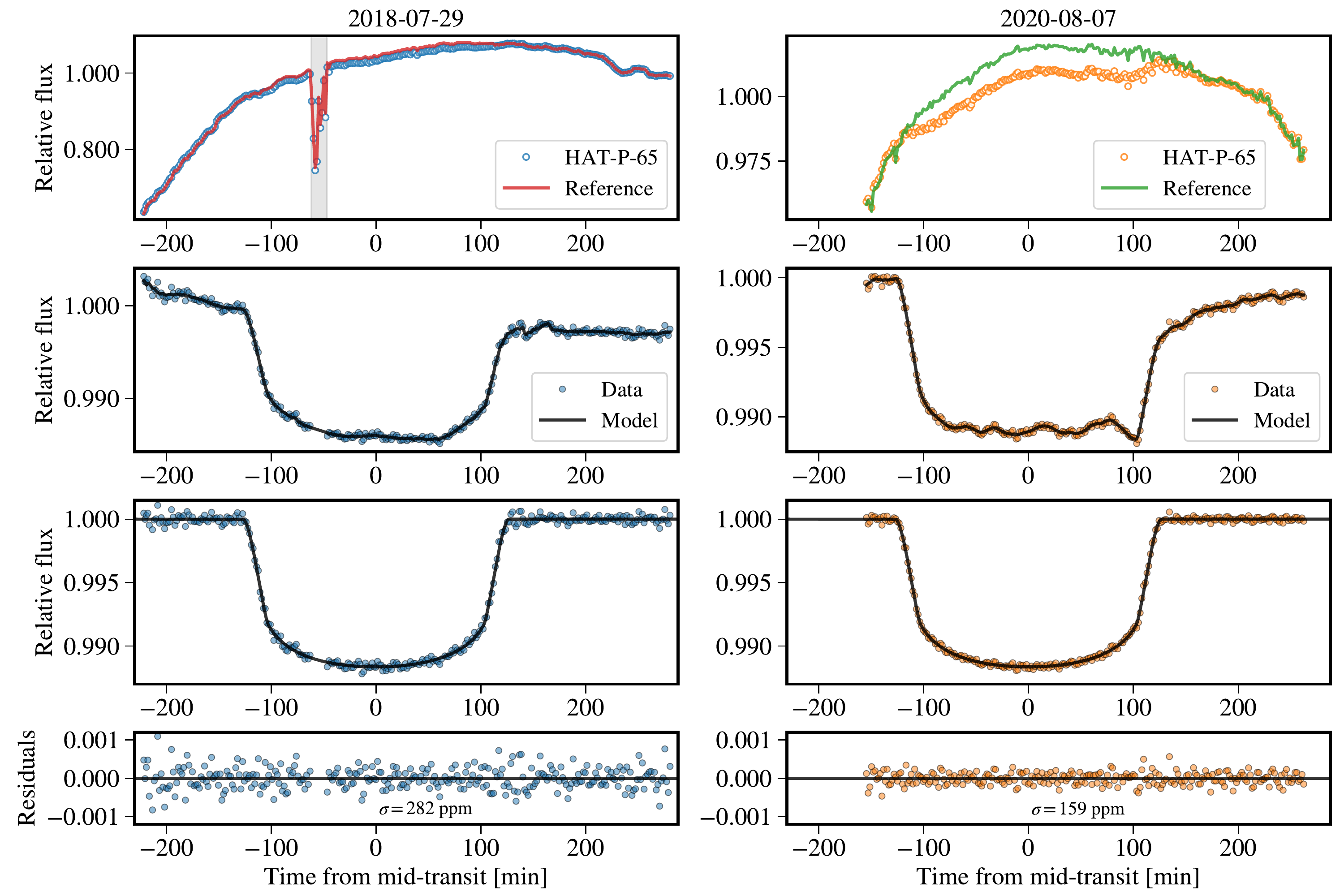}
\caption{White light curves of HAT-P-65b observed by GTC/OSIRIS on the nights of 2018 July 29 (left, Night 1) and 2020 August 7 (right, Night 2). From top to bottom are: i) raw flux time series of HAT-P-65 and its reference star, ii) raw white light curves of HAT-P-65 (i.e., normalized target-to-reference flux ratios), iii) white light curves corrected for systematics, iv) best-fit light-curve residuals. The best-fit models are shown in black.\label{fig:wlc}}
\end{figure*}

The white light curves are presented in Figure~\ref{fig:wlc}. The best-fit residuals have a standard deviation of 282~ppm (Night 1) and 159~ppm (Night 2), which are 2.94$\times$ and 1.63$\times$ photon noise, respectively. The derived parameters are presented in Table~\ref{tab:params}. Our transit parameters do not agree well with those reported in the discovery paper \citep[$i=84.2\pm 1.3$$^{\circ}$, $a/R_\star=4.57\pm 0.20$;][]{2016AJ....152..182H}. This can be probably attributed to the fact that all the follow-up observations in the discovery paper only covered partial transits. 

In addition to the joint analysis, we also performed the light-curve modeling for each transit individually. This resulted in $i=88.96^{+0.73}_{-0.99}$$^{\circ}$, $a/R_\star=5.188^{+0.034}_{-0.059}$, $R_\mathrm{p}/R_\star=0.1029^{+0.0035}_{-0.0036}$ for Night 1, and $88.89^{+0.78}_{-1.03}$$^{\circ}$, $5.234^{+0.036}_{-0.069}$, $0.0966^{+0.0034}_{-0.0035}$ for Night 2. While the best-fit transit depths are discrepant to some extent, the large uncertainties still make them consistent at 1.3$\sigma$. We note that the strong (but achromatic) systematics in Night 2 might have contributed to this bias.

\begin{figure*}
\centering
\includegraphics[width=\textwidth]{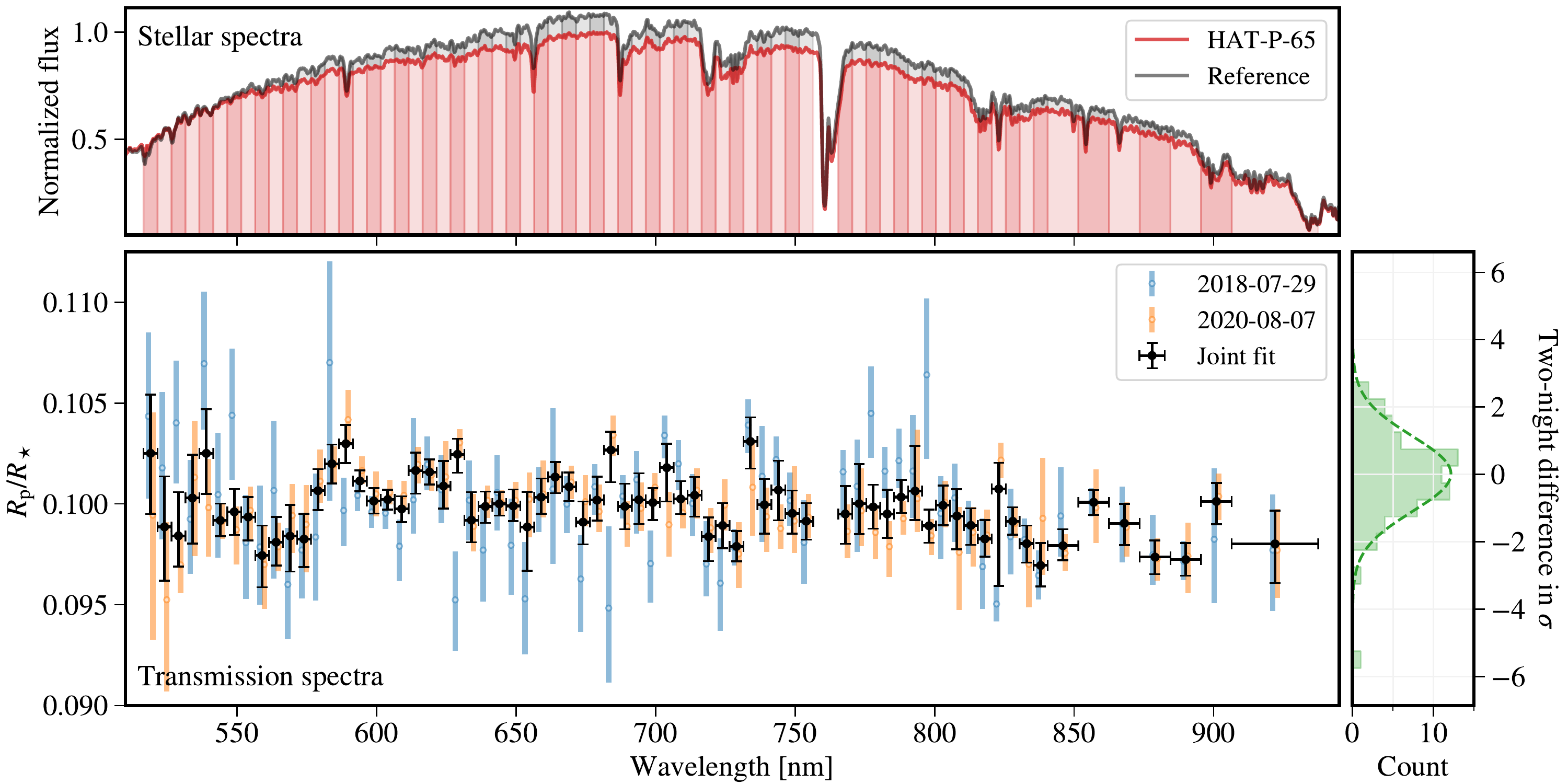}
\caption{Transmission spectra of HAT-P-65b measured jointly (black) and individually (blue and orange) for the two nights. The top sub-panel shows an example of the stellar spectra for HAT-P-65 and its reference star, along with the passbands used in this work marked in shaded colors. The right sub-panel presents the distribution histogram of the two-night differences, which has been normalized by the measurement uncertainties.\label{fig:ts_cmp}}
\end{figure*}

\begin{figure*}
\centering
\includegraphics[width=\textwidth]{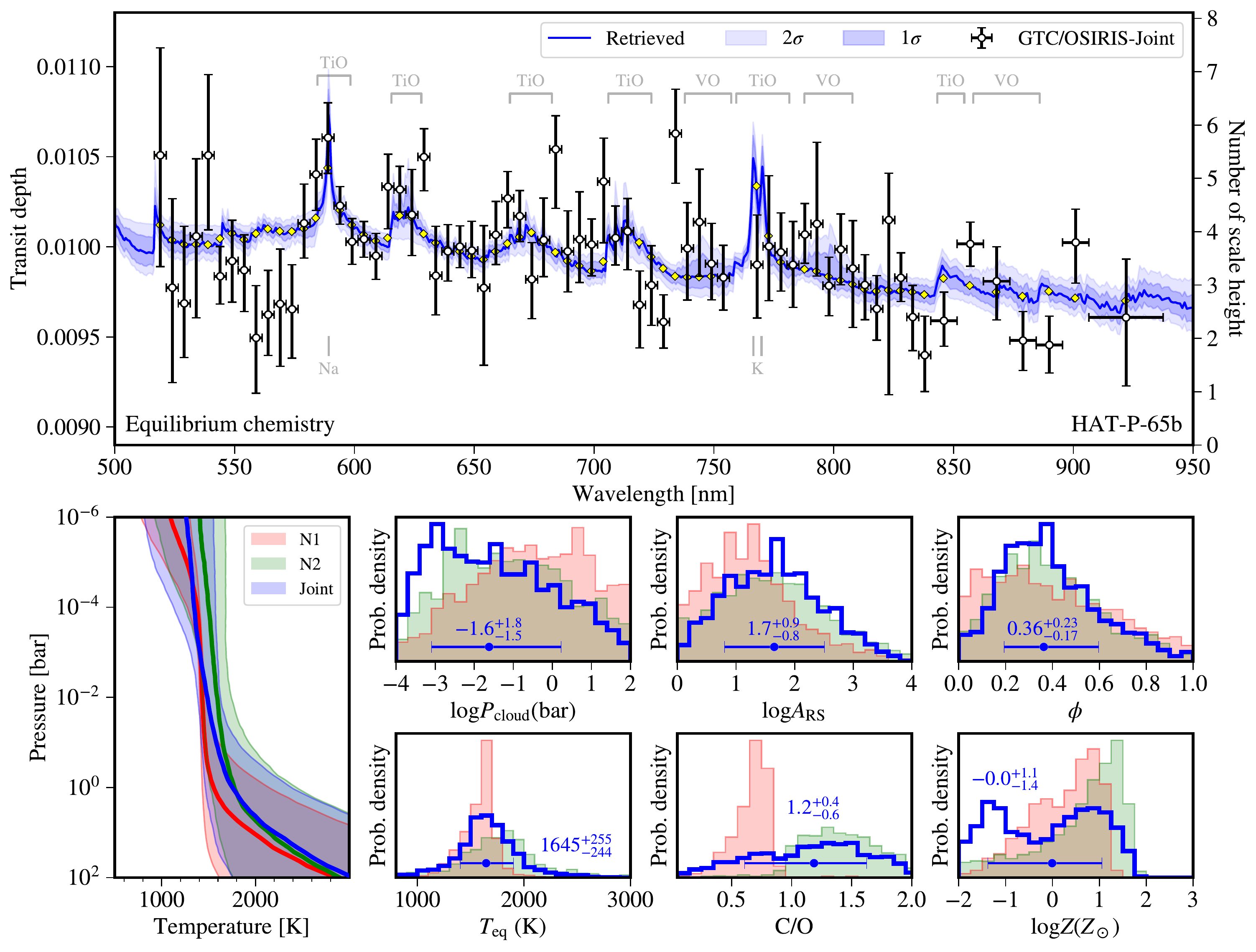}
\caption{Transmission spectrum of HAT-P-65b and retrieved atmospheric properties assuming equilibrium chemistry. The first row presents the jointly derived transmission spectrum (white circles) and retrieved models (blue line and shaded areas). The second and third rows present the retrieved temperature-pressure (T-P) profile, clout-top pressure $P_\mathrm{cloud}$, enhancement over H$_2$ Rayleigh scattering $A_\mathrm{RS}$, cloud coverage $\phi$, equilibrium temperature $T_\mathrm{eq}$ used in the T-P profile, C/O ratio, and atmospheric metallicity $Z$. The blue, red, and green lines and shaded areas refer to the retrieval results based on the joint, Night 1 (N1), and Night 2 (N2) transmission spectra, respectively. \label{fig:ts_retrieved_equ}}
\end{figure*}

\begin{figure*}
\centering
\includegraphics[width=\textwidth]{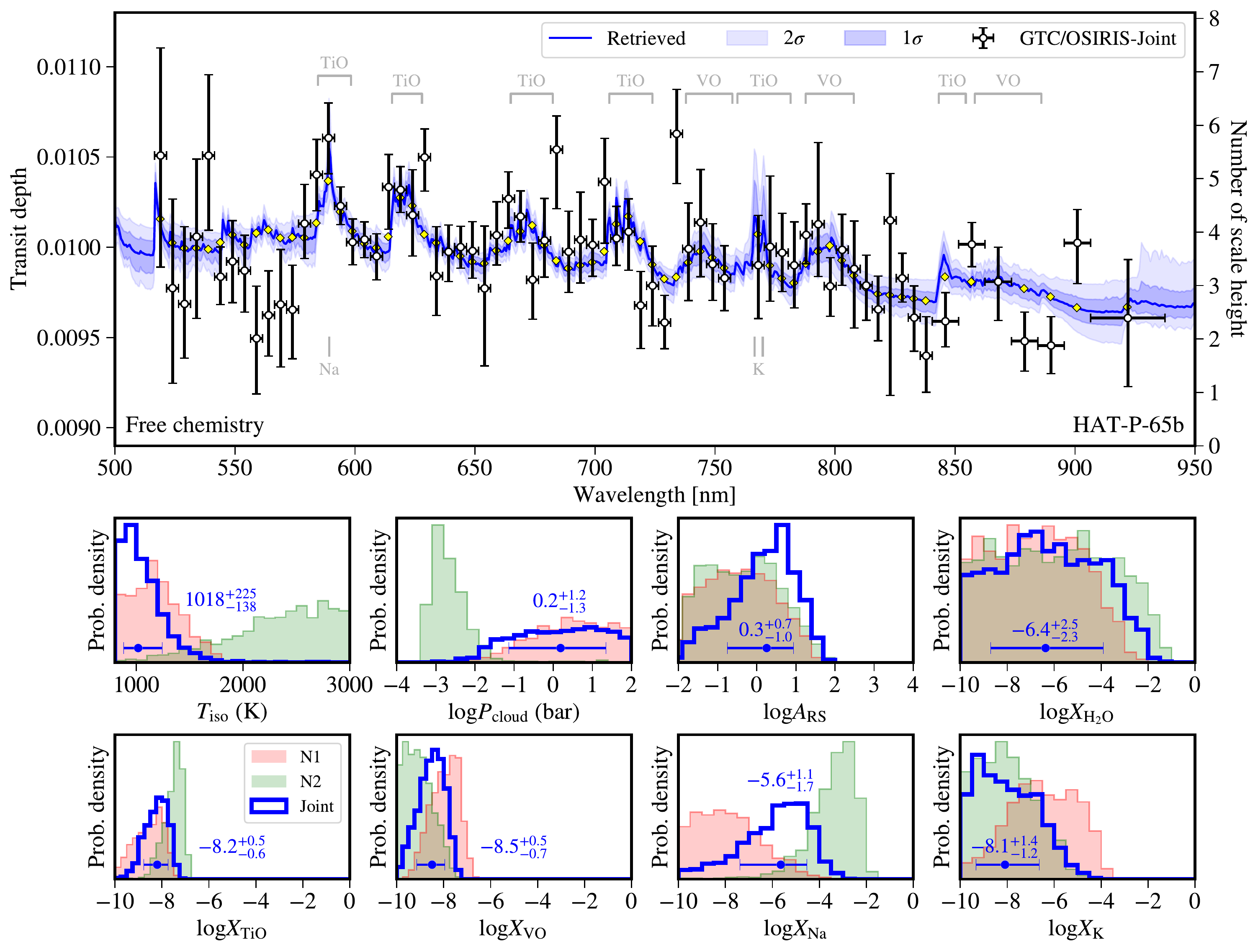}
\caption{Transmission spectrum of HAT-P-65b and retrieved atmospheric properties assuming free chemistry. The first row presents the jointly derived transmission spectrum (white circles) and retrieved models (blue line and shaded areas). The second and third rows present the retrieved isothermal temperature $T_\mathrm{iso}$, clout-top pressure $P_\mathrm{cloud}$, enhancement over H$_2$ Rayleigh scattering $A_\mathrm{RS}$, and mass fractions $X_i$ for H$_2$O, TiO, VO, Na, and K. The blue, red, and green lines and shaded areas refer to the retrieval results based on the joint, Night 1 (N1), and Night 2 (N2) transmission spectra, respectively. \label{fig:ts_retrieved_free}}
\end{figure*}

\subsection{Spectroscopic light curves}

Before modeling the spectroscopic light curves, we first derived the empirical common-mode systematics models by dividing the white light curves by the best-fit transit model. For each night, the spectroscopic light curves were divided by the corresponding empirical models to correct for the common-mode systematics. The corrected spectroscopic light curves were then fitted in a similar way as described in the last subsection, except that the values of $i$, $a/R_\star$, and $T_\mathrm{mid}$ were fixed to the best-fit values obtained for the white light curves. Fixing the values of $i$ and $a/R_\star$ to \citet{2016AJ....152..182H}'s has negligible impact on the derived transmission spectrum (see Appendix \ref{sec:hartman}). We chose the transit model as the GP mean function and time $t$ as the GP covariance matrix input (see Appendix \ref{sec:modelselect}). Consequently, the free parameters were ($R_\mathrm{p}/R_\star$, $u_1$, $u_2$, $\sigma_w$, $A$, $\tau_t$).

The spectroscopic light curves of the two nights were fitted in two separate runs. In the first run, the two nights were fitted individually, so as to examine the consistency between them. In the second run, the two nights were fitted jointly, sharing the same values for ($R_\mathrm{p}/R_\star$, $u_1$, $u_2$). In general, the spectroscopic light curves of Night 2 have better precision than Night 1. For Night 1, the standard deviation of the best-fit residuals achieve 1.3--2.1$\times$ photon noise, while for Night 2 it is 1.0--1.8$\times$ photon noise.

Figure \ref{fig:ts_cmp} gives an illustration of the spectroscopic passbands, and presents the individually and jointly fitted transmission spectra. The two nights are consistent with each other at 1.5$\sigma$ confidence level ($\chi^2=81.7$ for 68 degrees of freedom, hereafter dof), after excluding two outlier channels located at 629~nm and 823~nm in which the two nights are different for more than 3$\sigma$. Nevertheless, the histogram of the two-night differences follows a nearly Gaussian distribution. The derived transmission spectra are given in Table \ref{tab:ts} in Appendix \ref{sec:additional}.

\section{Transmission spectrum}
\label{sec:transpec}

The jointly fitted transmission spectrum of HAT-P-65b roughly spans five scale heights ($H/R_\star=0.001475$, where $H=k_\mathrm{B}T_\mathrm{eq}/\mu g_\mathrm{p}$). Two broad spectral features at $\sim$590~nm and $\sim$620~nm appear prominent. 

To calculate the Bayesian evidence $\mathcal{Z}$ for the spectral retrieval analyses, we used the Python package \texttt{PyMultiNest} \citep{2014A&A...564A.125B}, which relies on the MultiNest library \citep{2009MNRAS.398.1601F} and implements the multimodal nested sampling algorithm. For model comparison, we calculated the Bayes factor ($B_\mathrm{10}=\mathcal{Z}_1/\mathcal{Z}_0$) and adopted the criteria of $\ln B_{10}=1.0, 2.5, 5.0$ as the starting points of ``weak'', ``moderate'', ``strong'' evidence in favor of model 1 over model 0 \citep{2008ConPh..49...71T}, respectively. To compare with literature, the Bayes factor was converted to the traditional frequentist significance \citep{2008ConPh..49...71T,2013ApJ...778..153B}, but the number of ``sigmas'' does not necessarily mean a detection \citep{2021arXiv210308600W}.

We first compared HAT-P-65b's transmission spectrum to a flat line and a sloped line. The spectrum is inconsistent with a flat line at 4.2$\sigma$ level ($\chi^2=127.7$ for 69 dof). The flat and sloped lines have $\ln\mathcal{Z}=456.23$ and 466.20, respectively, indicating that the sloped line is preferred over the flat line at 4.8$\sigma$ level. The derived slope $\mathrm{d}(R_\mathrm{p}/R_\star)/\mathrm{d}(\ln\lambda)=-0.00405\pm 0.00078$ corresponds to a scattering index of $\alpha=-2.75\pm 0.53$ if it is assumed to be induced by a power law scattering cross section $\kappa=\kappa_0(\lambda/\lambda_0)^\alpha$ \citep{2008A&A...481L..83L} in an atmosphere at the equilibrium temperature. 

We then performed spectral retrieval analyses on the transmission spectrum using the Python packages \texttt{petitRADTRANS} \citep{2019arXiv190411504M} and \texttt{PyMultiNest}. We adopted the modified \citet{2010A&A...520A..27G} temperature-pressure (T-P) profile \citep{2019arXiv190411504M}, which consists of six free parameters. Following \citet{2017MNRAS.469.1979M}, the atmosphere was assumed to be covered by clouds and hazes at a fraction of ${\phi}$, with the rest being clear. The clouds and hazes were parameterized as a cloud-top pressure ($P_\mathrm{cloud}$) and an enhancement factor over nominal Rayleigh scattering ($A_\mathrm{RS}$). The reference pressure $P_0$ at the planet radius $R_\mathrm{p}=1.89$~$R_\mathrm{J}$ was a free parameter. We started from equilibrium chemistry, using two free parameters (C/O and metallicity $\log Z$) to interpolate mass fractions of H$_2$, He, CO, H$_2$O, HCN, C$_2$H$_2$, CH$_4$, PH$_3$, CO$_2$, NH$_3$, H$_2$S, VO, TiO, Na, K, SiO, e$^-$, H$^-$, H, and FeH in a pre-calculated chemical grid \citep{2017A&A...600A..10M}. Collision-induced absorption of H$_2$-H$_2$ and H$_2$-He and Rayleigh scattering of H$_2$ and He were also included. 

Figure \ref{fig:ts_retrieved_equ} presents the best retrieved model ($\ln\mathcal{Z}=475.79$, $\chi^2=77.9$ for 58 dof) assuming equilibrium chemistry, along with the posterior distributions of the T-P profile and atmospheric properties. This physics-motivated model is strongly favored over the flat-line (6.6$\sigma$) and sloped-line (4.8$\sigma$) models. We note that the data points at 629~nm and 734~nm contribute significantly to the resulting chi-square ($\Delta\chi^2=13.4$). We obtained a T-P profile that is nearly isothermal across a wide range of pressure levels. We retrieved an equilibrium temperature of $1645^{+255}_{-244}$~K, and a cloud coverage of $36^{+23}_{-17}$\%, which has a poorly constrained cloud-top pressure (0.8--1585~mbar) and a haze scattering amplitude that is 8--398$\times$ H$_2$ Rayleigh scattering. The C/O ratio and metallicity are not well constrained by current data.

We also performed the spectral retrieval analyses assuming free chemistry to search for the species responsible for the observed spectral signatures. In this case, we adopted an isothermal T-P profile and described the clouds/hazes properties using $P_\mathrm{cloud}$ and $A_\mathrm{RS}$, without the consideration of cloudy/clear sections. We only included H$_2$, He, TiO, VO, Na, K, H$_2$O in the full model, and set the mass fractions of the latter five as free parameters. Figure \ref{fig:ts_retrieved_free} shows the best retrieved model ($\ln\mathcal{Z}=476.34$, $\chi^2=76.6$ for 61 dof; $\Delta\chi^2=14.4$ from 629~nm and 734~nm) and posterior distributions of atmospheric properties. The mass fraction posterior distributions show clear modes for TiO, VO, Na at $\log X=-8.2^{+0.5}_{-0.6}$, $-8.5^{+0.5}_{-0.7}$, $-5.6^{+1.1}_{-1.7}$, but not for K and H$_2$O. We experimented with decreasing the prior limit and found that the TiO posterior remained the same while the posteriors of VO and Na exhibited tails bound by the lower prior limit. On the other hand, we removed certain species one by one and compared their Bayesian evidence to that of the full model. Removing any one of TiO, VO, Na would decrease the Bayesian evidence by $\Delta\ln\mathcal{Z}=-5.2, -0.9, -1.2$, respectively. This indicates that the presence of TiO is strongly favored.

Since Night 2 is heavily weighted in the joint analysis due to higher precision, we performed the retrieval analyses in each night to inspect their individual constraints on the atmospheric properties. The derived parameters and statistics for all the analyses are presented in Tables \ref{tab:retrievalparam} and \ref{tab:retrievalstat} in Appendix \ref{sec:additional}. We found weak-to-moderate evidence for TiO in both nights and the measured mass fractions were in broad agreement. In contrast, we only found moderate evidence for VO in Night 1 and moderate evidence for Na in Night 2, but no evidence for them in the other night. This is consistent with the results obtained in the joint transmission spectrum, highlighting the importance to conduct repeated observations to improve the credibility of any detections. 

Finally, we investigated the impact of stellar heterogeneity by adding contamination of spots and faculae to the free-chemistry retrieval analyses. We included the component of stellar contamination in a way similar to \citet{2021MNRAS.500.5420C}. For both individual and joint transmission spectra, the Bayesian evidence decreases significantly if no planetary atmosphere is considered, indicating that the spots/faculae alone cannot explain our observations. The retrieved planetary atmospheric properties remain consistent with those obtained in the case when no spots/faculae are included (see Figure \ref{fig:ts_retrieved_spot} in Appendix \ref{sec:additional}), except for Na mass fraction, which increases from $\log X=-5.6^{+1.1}_{-1.7}$  to $-2.5^{+1.3}_{-1.5}$. On the other hand, the Bayes factors disfavor stellar contaminations in the retrievals for Night 1 and joint transmission spectra, but not for Night 2. However, we confirm that even if stellar contamination exists in Night 2, it does not contribute any spectral signatures mimicking TiO in the case of HAT-P-65 (see Figure \ref{fig:stellar_contam} in Appendix \ref{sec:additional}).

\section{Summary and discussion}
\label{sec:discuss}

We observed two transits of HAT-P-65b using the OSIRIS spectrograph on the 10.4~m GTC, and derived two consistent individual transmission spectra. We then jointly fitted the two nights to derive the final transmission spectrum, in which at least two of the TiO absorption bands were clearly resolved (i.e., $\sim$585--598~nm and $\sim$615--628~nm). We performed spectral retrieval analyses on the joint transmission spectrum and found a relatively clear atmosphere showing strong evidence for TiO. The analyses on the individual transmission spectra instead reveal weak-to-moderate evidence for TiO in both nights, but only moderate evidence for VO or Na in one of the nights (no evidence in the other).

The detection of TiO in transmission spectroscopy is still rare and intriguing. Several mechanisms have been proposed to explain the lack of TiO detection, such as cold trapping either in deeper atmosphere or on cooler nightside \citep{2003ApJ...594.1011H,2009ApJ...699.1487S,2013A&A...558A..91P}, photodissociation \citep{2010ApJ...720.1569K}, or thermal dissociation \citep{2018A&A...617A.110P,2018ApJ...866...27L}. The strong evidence for TiO suggests that these mechanisms have not completely removed TiO from HAT-P-65b's observable atmosphere. 

The presence of TiO in the upper atmosphere could absorb incoming stellar radiation and introduce a thermal inversion in the T-P profile \citep{2003ApJ...594.1011H,2008ApJ...678.1419F}. We have retrieved a nearly isothermal T-P profile for the atmosphere at the day-night terminator. Given the equilibrium temperature of $\sim$1930~K, it is likely that HAT-P-65b's dayside temperature is not sufficiently hot to thermally dissociate TiO. Therefore, it is possible to look for the TiO signature in follow-up secondary-eclipse observations and to investigate its role in changing the vertical thermal structure.

We note that it is difficult to claim the detection of either TiO, VO, or Na at this stage. In particular, the Na line overlaps with the TiO band at $\sim$585--598~nm, which are difficult for low-resolution transmission spectroscopy to resolve. Fortunately, high-resolution Doppler spectroscopy, acquired using state-of-the-art ultrastable spectrographs like ESPRESSO \citep[e.g.,][]{2020Natur.580..597E,2020A&A...635A.171C,2021arXiv210104094C}, will likely be able to unambiguously distinguish different atomic and molecular species and possibly their morning-evening differences, which will strongly improve our understanding of atmospheric circulation in hot Jupiters.

\acknowledgments

    G.\,C. acknowledges the support by the B-type Strategic Priority Program of the Chinese Academy of Sciences (Grant No.\,XDB41000000), the National Natural Science Foundation of China (Grant No. 42075122), the Natural Science Foundation of Jiangsu Province (Grant No.\,BK20190110), Youth Innovation Promotion Association CAS (2021315), and the Minor Planet Foundation of the Purple Mountain Observatory. 
    This work is partly financed by the Spanish Ministry of Economics and Competitiveness through grant ESP2013-48391-C4-2-R. 
    This work is based on observations made with the Gran Telescopio Canarias (GTC), installed at the Spanish Observatorio del Roque de los Muchachos of the Instituto de Astrof\'{i}sica de Canarias, in the island of La Palma.
    This work has made use of the VizieR catalog access tool, CDS, Strasbourg, France \citep{2000A&AS..143...23O}. 
    The authors thank the anonymous referee for their constructive comments on the manuscript.

%

\facilities{GTC(OSIRIS)}


\software{
\texttt{Matplotlib} \citep{2007CSE.....9...90H},
\texttt{batman} \citep{2015PASP..127.1161K},
\texttt{george} \citep{2015ITPAM..38..252A}
\texttt{emcee} \citep{2013PASP..125..306F},
\texttt{petitRADTRANS} \citep{2019arXiv190411504M}, 
\texttt{PyMultiNest} \citep{2014A&A...564A.125B}
}



\appendix

\section{Impact of flux dilution by a companion star. \label{sec:dilution}}

\begin{figure}[h!]
\centering
\includegraphics[width=0.6\textwidth]{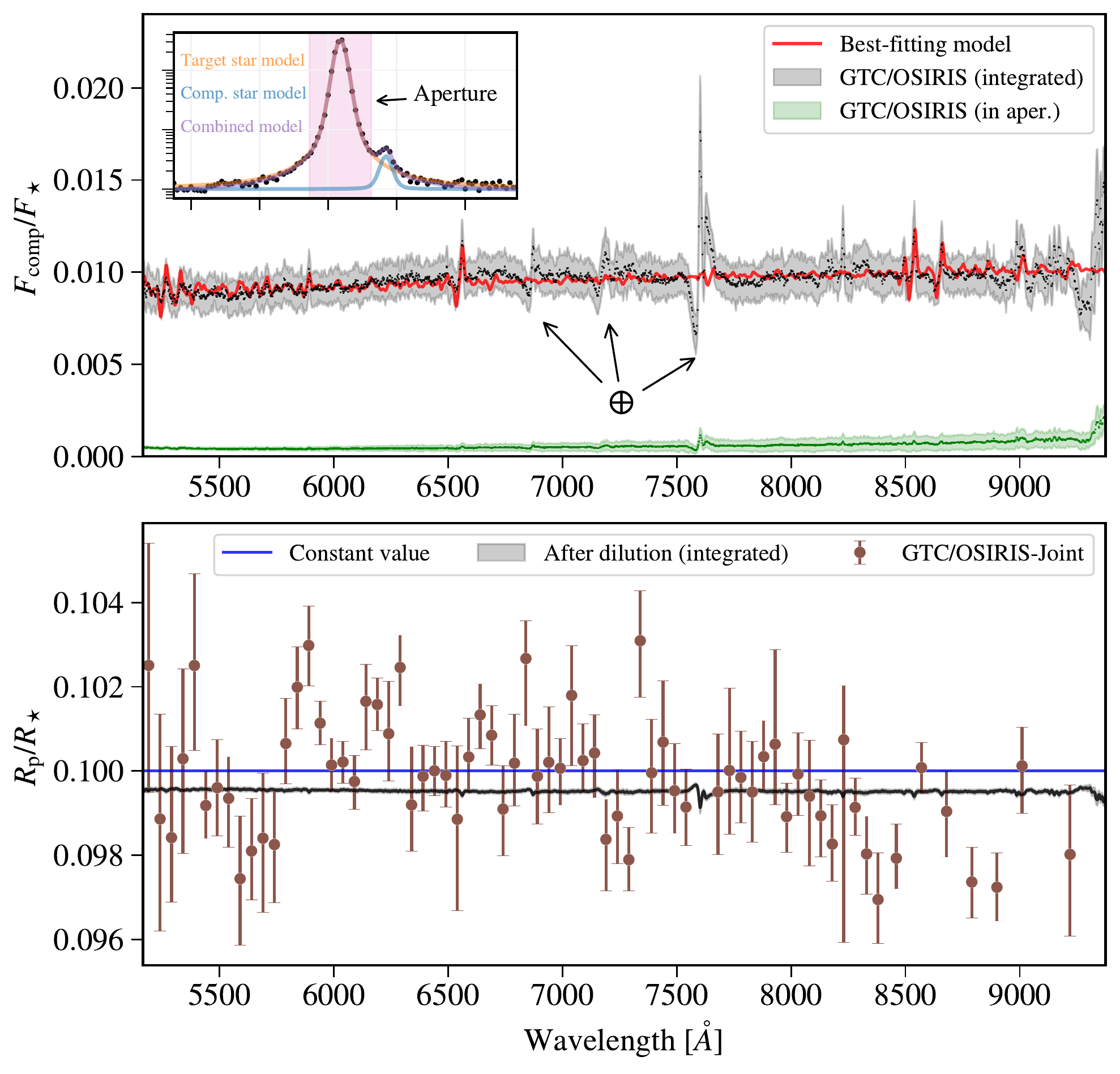}
\caption{{\it Top:} fully-integrated (black) and in-aperture companion-to-target flux-ratio spectra. The red line shows the best-fit flux-ratio model. The inset illustrates the PSF fitting for HAT-P-65 and its companion star. The pink area indicates the aperture adopted in Night 1. {\it Bottom:} dilution effect compared to joint transmission spectrum. The blue line shows a constant value (0.1000), and the black line and shaded areas correspond to the values after being diluted by fully-integrated flux-ratio spectrum. \label{fig:dilution}}
\end{figure}

\citet{2016AJ....152..182H} resolved a background companion star to HAT-P-65 with a similar effective temperature at a distance of 3.6$''$ ($\Delta J=4.91\pm0.01$ mag and $\Delta K=4.95\pm0.03$). This companion star is also resolved in our GTC/OSIRIS observation, with a projected distance of 13.13 pixels (3.34$''$) along the slit. We adopted an aperture radius of 9 pixels (2.29$''$) and 8 pixels (2.03$''$) in the spectral extraction. To quantitatively assess the impact of the dilution of the companion star, we fitted the PSF of HAT-P-65 and its companion following the methodology described in \citet{2021MNRAS.500.5420C}. We used the out-of-transit spectra images to calculate the fully integrated companion-to-target flux ratio spectrum, and recorded the standard deviation as its uncertainties. Similarly, we also calculated the flux-ratio spectrum within the aperture that was used to extract the spectra of HAT-P-65. We fitted the integrated flux-ratio spectrum using the spectral templates from the PHOENIX stellar atmospheres \citep{2013A&A...553A...6H}. This resulted in the stellar parameters $T_\mathrm{eff}=5443^{+18}_{-14}$~K, $\log g=3.11^{+0.17}_{-0.11}$, $\mathrm{[Fe/H]}=-0.11^{+0.21}_{-0.08}$ for the companion star, with a flux rescaling factor of $f=0.01289^{+0.00020}_{-0.00034}$ and $T_\mathrm{eff}=5835$~K, $\log g=4.18$, $\mathrm{[Fe/H]}=0.10$ being fixed for HAT-P-65. In Figure \ref{fig:dilution}, we present the integrated and in-aperture flux-ratio measurements. The dilution caused by the in-aperture companion flux is almost negligible. Even if the companion were fully included in the aperture (i.e., the integrated spectrum), the wavelength-dependent dilution difference is too small to explain any spectral signatures observed in the transmission spectrum. Therefore, we conclude that this companion star does not impact our results.

\section{Model selection for light-curve analysis. \label{sec:modelselect}}

To determine the optimal GP mean function, we experimented with the transit model, the transit multiplied by a linear trend or by a quadratic trend. For the GP covariance matrix input, we also tested different combinations of state vectors $\hat{x}_\alpha$ (e.g., time sequence $t$, spectral drift $x$, spatial drift $y$, spatial FWHM $s_y$). We used \texttt{PyMultiNest} to implement the multimodal nested sampling algorithm and to calculate the natural log of the Bayesian evidence ($\ln\mathcal{Z}$). Table \ref{tab:selectmodel} lists the resulting $\ln\mathcal{Z}$ for all the tested models. For the white light curves, Model 2 gives the highest evidence and is significantly better than the other models ($>$3$\sigma$ for $\Delta\ln\mathcal{Z}>3.15$), while Model 1 is the best for the spectroscopic light curves that have been corrected for the common-mode systematics. We note that the derived transit parameters were in general consistent even if a different model was used.

\begin{deluxetable}{ccccccc}
\tablecaption{Model selection for light-curve analysis. \label{tab:selectmodel}}
\tablewidth{0pt}
\tabletypesize{\footnotesize}
\tablehead{
\colhead{\#} & 
\multicolumn{2}{c}{Model} & 
\multicolumn{2}{c}{White} &
\multicolumn{2}{c}{Spectroscopic} \\ 
\colhead{} & 
\colhead{GP} & 
\colhead{Trend} & 
\colhead{$\ln\mathcal{Z}$} & 
\colhead{$\Delta\ln\mathcal{Z}$} & 
\colhead{$\ln\mathcal{Z}$} & 
\colhead{$\Delta\ln\mathcal{Z}$}
} 
\startdata
    1 & GP($t$)     & -- & 3615.89 & $-$4.63 & 204343.95 & 0\\ \noalign{\smallskip}
    2 & GP($t,x$)   & -- & 3620.53 & 0       & 204294.21 & $-$49.7\\ \noalign{\smallskip}
    3 & GP($t,y$)   & -- & 3617.38 & $-$3.15 & 204337.49 & $-$6.5\\ \noalign{\smallskip}
    4 & GP($t,s_y$) & -- & 3616.49 & $-$4.03 & 204301.83 & $-$42.1\\ \noalign{\smallskip}
    5 & GP($t$)     & $B=c_0+c_1t$ & 3602.86 & $-$17.66 & 203044.76 & $-$1299.2\\ \noalign{\smallskip}
    6 & GP($t,x$)   & $B=c_0+c_1t$ & 3608.32 & $-$12.20 & 202898.40 & $-$1445.6\\ \noalign{\smallskip}
    7 & GP($t,y$)   & $B=c_0+c_1t$ & 3605.45 & $-$15.08 & 202968.37 & $-$1375.6\\ \noalign{\smallskip}
    8 & GP($t,s_y$) & $B=c_0+c_1t$ & 3605.83 & $-$14.69 & 202916.09 & $-$1427.9\\ \noalign{\smallskip}
    9 & GP($t$)     & $B=c_0+c_1t+c_2t^2$ & 3608.54 & $-$11.98 & 202745.46 & $-$1598.5\\ \noalign{\smallskip}
    10 & GP($t,x$)   & $B=c_0+c_1t+c_2t^2$ & 3608.83 & $-$11.69 & 202604.22 & $-$1739.7\\ \noalign{\smallskip}
    11 & GP($t,y$)   & $B=c_0+c_1t+c_2t^2$ & 3609.66 & $-$10.87 & 202652.96 & $-$1691.0\\ \noalign{\smallskip}
    12 & GP($t,s_y$) & $B=c_0+c_1t+c_2t^2$ & 3611.73 & $-$8.80  & 202618.98 & $-$1725.0\\ \noalign{\smallskip}
\enddata
\end{deluxetable}

\section{Transmission spectrum derived from different orbital parameters.\label{sec:hartman}}

\begin{figure}
\centering
\includegraphics[width=0.9\textwidth]{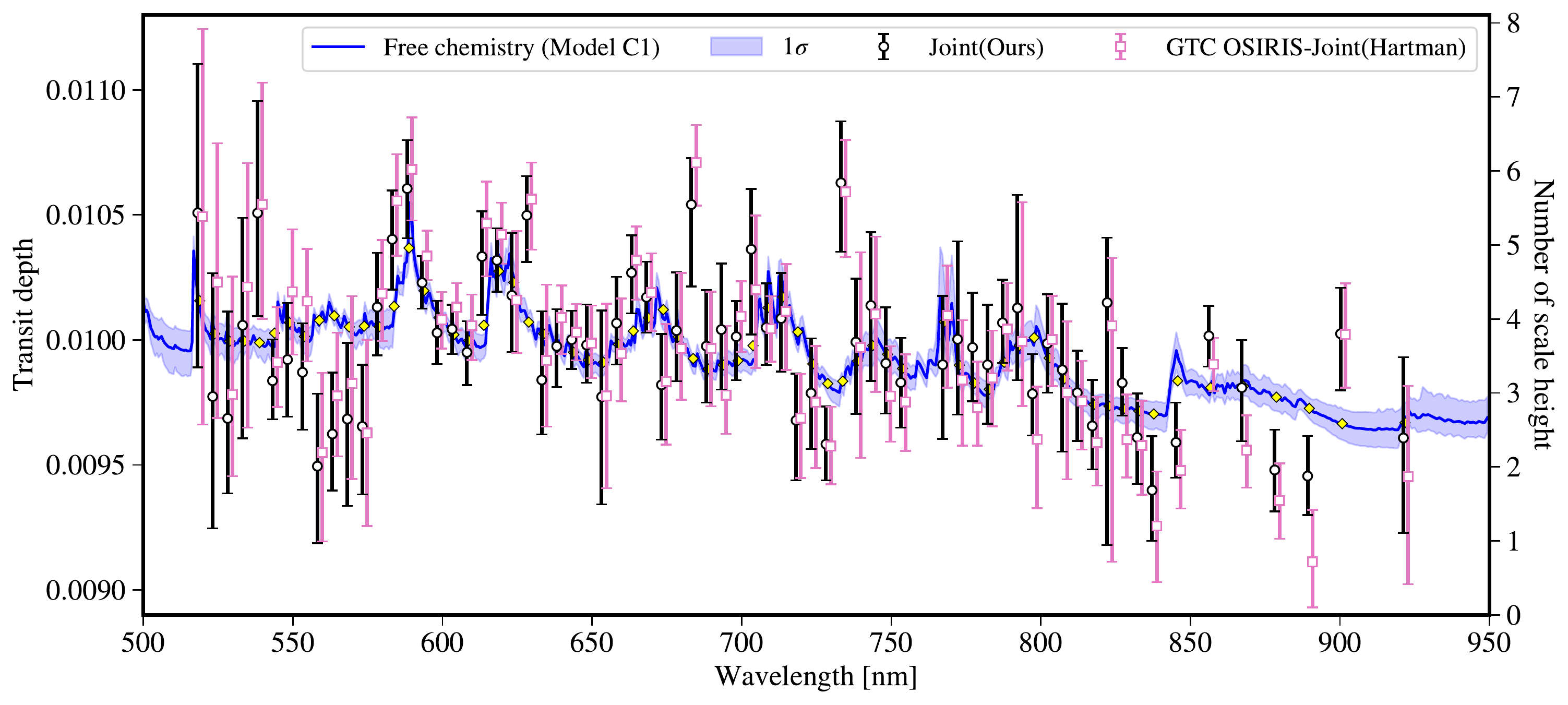}
\caption{Comparison of transmission spectra based on different orbital parameters. Our transmission spectrum (black cicles) is self-consistent, with $i=89.10$$^\circ$ and $a/R_\star=5.221$ determined from our white light curves. The other one (pink squares) is derived using $i=84.2$$^\circ$ and $a/R_\star=4.57$ \citep{2016AJ....152..182H}. The two transmission spectra agree well with each other. The model spectrum is the same as the one shown in Figure \ref{fig:ts_retrieved_free}.\label{fig:ts_hartman}}
\end{figure}

It has been shown that due to the limb-darkening effect, fixing impact parameter to imperfectly estimated values could introduce wavelength-dependent offsets to transmission spectrum in certain cases \citep{2018A&A...620A.142A,2020A&A...640A.134A}. We have derived the values for $i$ and $a/R_\star$ that are significantly different from the discovery paper \citep{2016AJ....152..182H}. To assess how this would impact our derived transmission spectrum, we performed the same analyses for the white and spectroscopic light curves as we did in Section \ref{sec:lc}, except that we always held $i=84.2$$^\circ$ and $a/R_\star=4.57$ fixed. Figure \ref{fig:ts_hartman} presents the derived transmission spectrum based on \citet{2016AJ....152..182H}'s orbital parameters, which is consistent with our self-consistently derived transmission spectrum. We conclude that the orbital parameters ($i$ and $a/R_\star$) are not the origin of the detected spectral signatures in our case.

\section{Additional tables and figures. \label{sec:additional}}

\begin{figure}
\centering
\includegraphics[width=\textwidth]{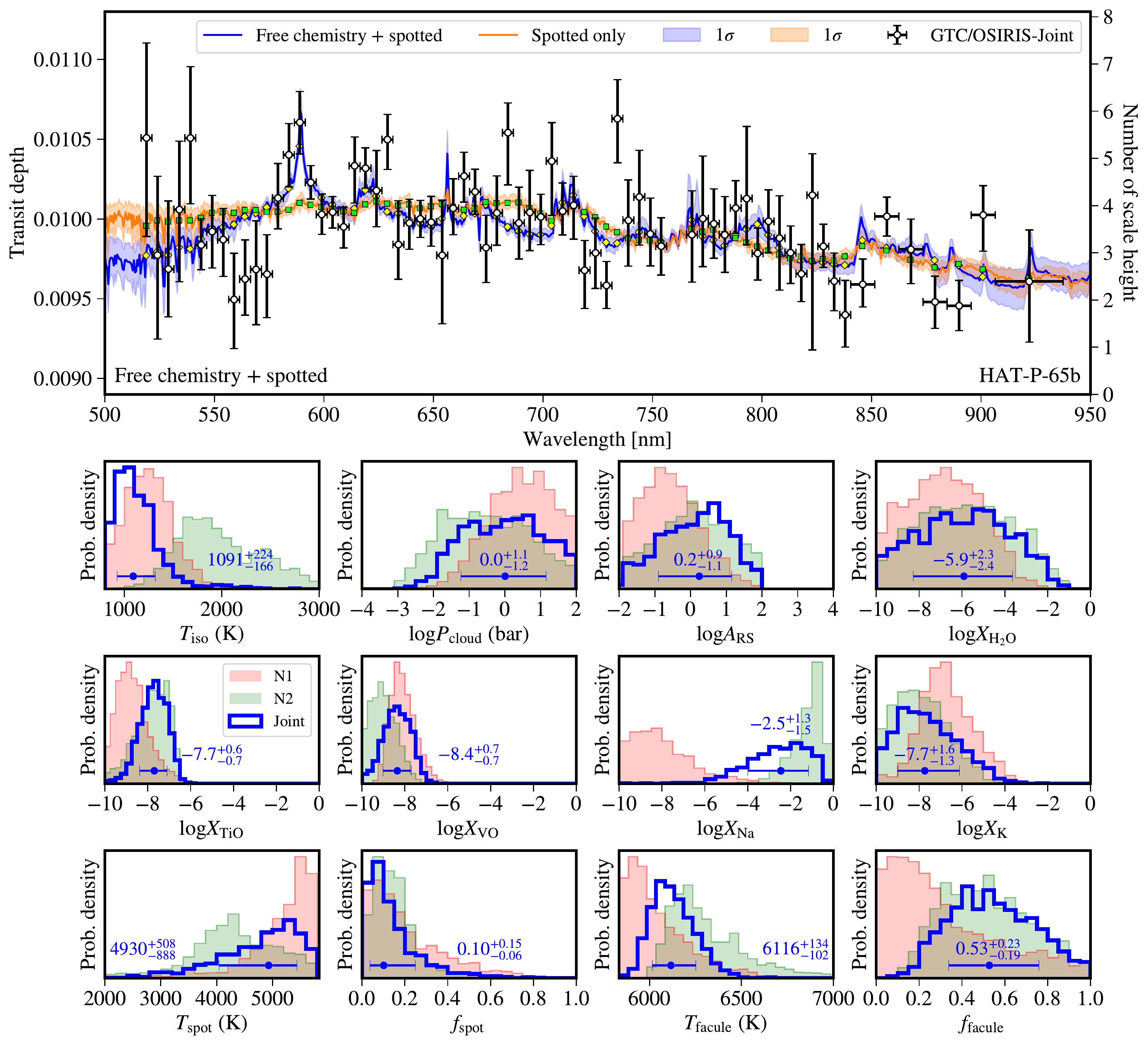}
\caption{Transmission spectrum of HAT-P-65b and retrieved atmospheric properties assuming free chemistry along with contamination from spots and faculae. The first row presents the jointly derived transmission spectrum (white circles) and retrieved models (blue line and shaded areas). For comparison, we show the best model from another retrieval analysis where only spots/faculae contamination is used to fit the data (orange line and shaded areas). The second to fourth rows present the retrieved planetary atmosphere properties (see the description in Figure \ref{fig:ts_retrieved_free}) and stellar spots/faculae properties (temperatures $T_\mathrm{spot}$, $T_\mathrm{faculae}$ and fractions $f_\mathrm{spot}$, $f_\mathrm{faculae}$). The blue, red, and green lines and shaded areas refer to the retrieval results based on the joint, Night 1 (N1), and Night 2 (N2) transmission spectra, respectively.\label{fig:ts_retrieved_spot}}
\end{figure}

\begin{figure}
\centering
\includegraphics[width=0.9\textwidth]{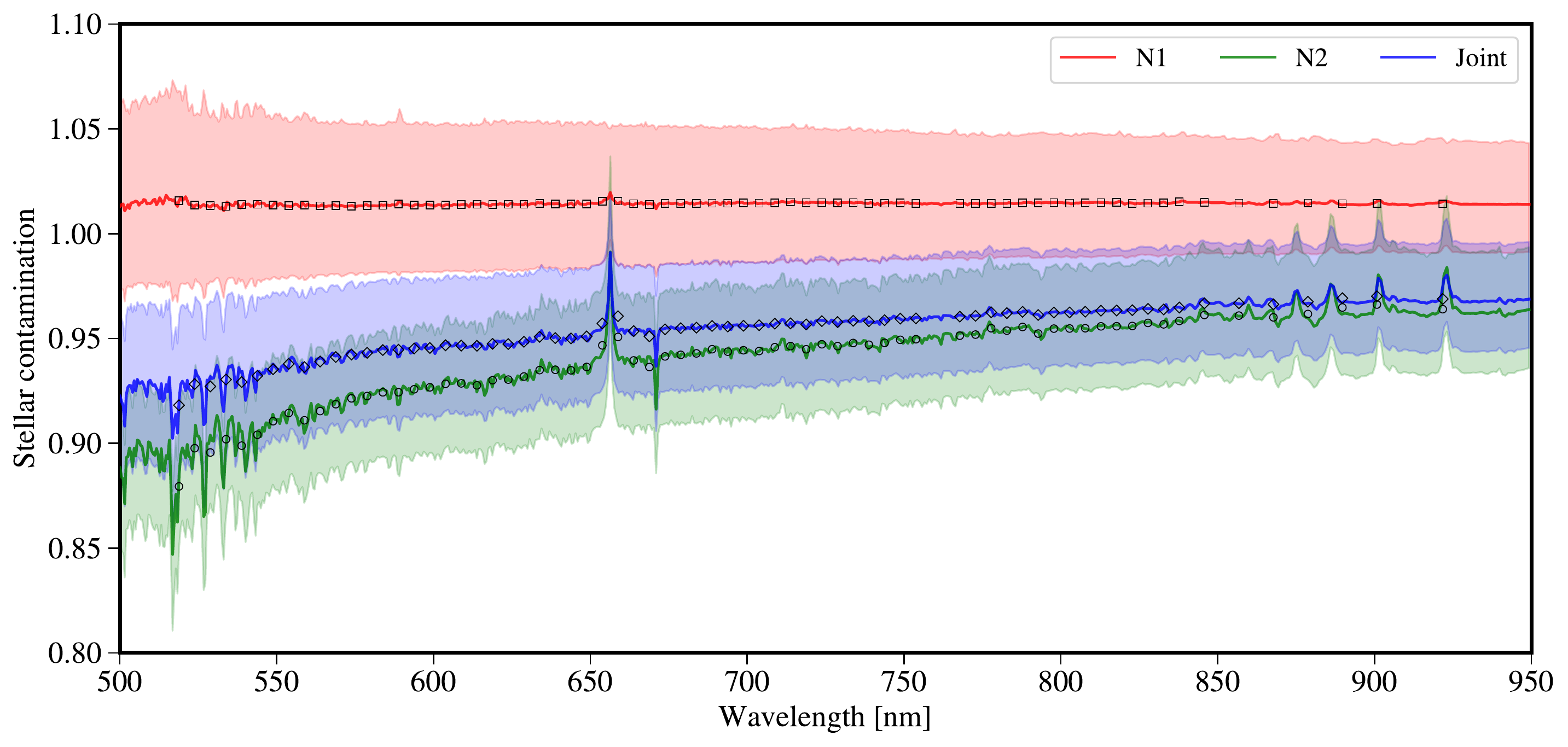}
\caption{Stellar contaminations obtained in the retrieval analyses assuming free-chemistry with spots/faculae (see Figure \ref{fig:ts_retrieved_spot}), for Night 1 (N1, red), Night 2 (N2, green), and joint (blue) transmission spectra, respectively. The spectral signatures observed at $\sim$585--598~nm and $\sim$615--628~nm in the transmission spectra cannot be explained by these contaminations. \label{fig:stellar_contam}}
\end{figure}

Table \ref{tab:ts} presents the transmission spectra derived from the individual and joint light-curve analyses. Tables \ref{tab:retrievalparam} and \ref{tab:retrievalstat} give the parameters and statistics obtained in the spectral retrieval analyses performed on these individual and joint transmission spectra. Figure \ref{fig:ts_retrieved_spot} shows the retrieved models and posterior distributions of planetary and stellar parameters assuming free-chemistry planetary atmosphere with stellar spots/faculae contamination. Figure \ref{fig:stellar_contam} shows the stellar contaminations obtained in the retrieval analyses where both free-chemistry planetary atmosphere and stellar spots/faculae contamination are considered (i.e., Model D1 in Table \ref{tab:retrievalstat}).

\begin{deluxetable}{cccccc}
\tablecaption{Transmission spectrum of HAT-P-65b.\label{tab:ts}}
\tablewidth{0pt}
\tabletypesize{\tiny}
\tablehead{
\colhead{$\lambda$ (nm)} & 
\colhead{$u_1$ prior} & 
\colhead{$u_2$ prior} &
\colhead{$R_\mathrm{p}/R_\star$ (N1)} &
\colhead{$R_\mathrm{p}/R_\star$ (N2)} &
\colhead{$R_\mathrm{p}/R_\star$ (Joint)}
} 
\startdata
    517--522 & $\mathcal{N}(0.526,0.064^2)$ & $\mathcal{N}(0.219,0.048^2)$ & $0.1043 ^{+0.0042}_{-0.0041}$ & $0.0994 ^{+0.0051}_{-0.0062}$ & $0.1025 ^{+0.0029}_{-0.0030}$\\
    522--526 & $\mathcal{N}(0.512,0.060^2)$ & $\mathcal{N}(0.237,0.040^2)$ & $0.1018 ^{+0.0038}_{-0.0036}$ & $0.0952 ^{+0.0039}_{-0.0045}$ & $0.0989 ^{+0.0025}_{-0.0027}$\\
    527--532 & $\mathcal{N}(0.499,0.057^2)$ & $\mathcal{N}(0.244,0.037^2)$ & $0.1040 ^{+0.0031}_{-0.0030}$ & $0.0970 ^{+0.0015}_{-0.0014}$ & $0.0984 ^{+0.0022}_{-0.0015}$\\
    532--536 & $\mathcal{N}(0.491,0.058^2)$ & $\mathcal{N}(0.247,0.038^2)$ & $0.0992 ^{+0.0029}_{-0.0027}$ & $0.1013 ^{+0.0028}_{-0.0039}$ & $0.1003 ^{+0.0021}_{-0.0023}$\\
    537--542 & $\mathcal{N}(0.485,0.058^2)$ & $\mathcal{N}(0.251,0.038^2)$ & $0.1070 ^{+0.0036}_{-0.0033}$ & $0.0998 ^{+0.0024}_{-0.0024}$ & $0.1025 ^{+0.0022}_{-0.0020}$\\
    542--546 & $\mathcal{N}(0.478,0.059^2)$ & $\mathcal{N}(0.256,0.039^2)$ & $0.1005 ^{+0.0031}_{-0.0031}$ & $0.0991 ^{+0.0009}_{-0.0009}$ & $0.0992 ^{+0.0008}_{-0.0008}$\\
    547--552 & $\mathcal{N}(0.477,0.056^2)$ & $\mathcal{N}(0.252,0.037^2)$ & $0.1044 ^{+0.0033}_{-0.0032}$ & $0.0987 ^{+0.0011}_{-0.0017}$ & $0.0996 ^{+0.0011}_{-0.0011}$\\
    552--556 & $\mathcal{N}(0.467,0.056^2)$ & $\mathcal{N}(0.258,0.036^2)$ & $0.0981 ^{+0.0024}_{-0.0028}$ & $0.0997 ^{+0.0012}_{-0.0015}$ & $0.0994 ^{+0.0010}_{-0.0012}$\\
    557--562 & $\mathcal{N}(0.458,0.056^2)$ & $\mathcal{N}(0.261,0.036^2)$ & $0.0979 ^{+0.0030}_{-0.0029}$ & $0.0970 ^{+0.0019}_{-0.0022}$ & $0.0974 ^{+0.0015}_{-0.0016}$\\
    562--566 & $\mathcal{N}(0.456,0.055^2)$ & $\mathcal{N}(0.261,0.035^2)$ & $0.1007 ^{+0.0035}_{-0.0044}$ & $0.0979 ^{+0.0012}_{-0.0012}$ & $0.0981 ^{+0.0012}_{-0.0012}$\\
    566--571 & $\mathcal{N}(0.449,0.052^2)$ & $\mathcal{N}(0.263,0.032^2)$ & $0.0960 ^{+0.0028}_{-0.0027}$ & $0.0994 ^{+0.0016}_{-0.0018}$ & $0.0984 ^{+0.0015}_{-0.0018}$\\
    572--576 & $\mathcal{N}(0.441,0.054^2)$ & $\mathcal{N}(0.271,0.033^2)$ & $0.0977 ^{+0.0017}_{-0.0024}$ & $0.0990 ^{+0.0019}_{-0.0024}$ & $0.0983 ^{+0.0013}_{-0.0014}$\\
    576--581 & $\mathcal{N}(0.439,0.053^2)$ & $\mathcal{N}(0.268,0.033^2)$ & $0.0984 ^{+0.0031}_{-0.0032}$ & $0.1011 ^{+0.0016}_{-0.0011}$ & $0.1007 ^{+0.0011}_{-0.0010}$\\
    582--586 & $\mathcal{N}(0.428,0.053^2)$ & $\mathcal{N}(0.276,0.032^2)$ & $0.1070 ^{+0.0050}_{-0.0068}$ & $0.1019 ^{+0.0009}_{-0.0010}$ & $0.1020 ^{+0.0010}_{-0.0010}$\\
    586--591 & $\mathcal{N}(0.428,0.055^2)$ & $\mathcal{N}(0.269,0.035^2)$ & $0.0997 ^{+0.0016}_{-0.0018}$ & $0.1042 ^{+0.0015}_{-0.0011}$ & $0.1030 ^{+0.0009}_{-0.0010}$\\
    592--596 & $\mathcal{N}(0.420,0.052^2)$ & $\mathcal{N}(0.277,0.032^2)$ & $0.1004 ^{+0.0008}_{-0.0008}$ & $0.1017 ^{+0.0007}_{-0.0007}$ & $0.1011 ^{+0.0005}_{-0.0005}$\\
    596--601 & $\mathcal{N}(0.413,0.052^2)$ & $\mathcal{N}(0.278,0.031^2)$ & $0.0996 ^{+0.0008}_{-0.0008}$ & $0.1005 ^{+0.0011}_{-0.0011}$ & $0.1001 ^{+0.0006}_{-0.0006}$\\
    602--606 & $\mathcal{N}(0.407,0.051^2)$ & $\mathcal{N}(0.281,0.030^2)$ & $0.0995 ^{+0.0007}_{-0.0008}$ & $0.1006 ^{+0.0006}_{-0.0006}$ & $0.1002 ^{+0.0005}_{-0.0005}$\\
    606--611 & $\mathcal{N}(0.407,0.050^2)$ & $\mathcal{N}(0.277,0.030^2)$ & $0.0979 ^{+0.0013}_{-0.0017}$ & $0.1004 ^{+0.0007}_{-0.0007}$ & $0.0998 ^{+0.0006}_{-0.0007}$\\
    612--616 & $\mathcal{N}(0.401,0.050^2)$ & $\mathcal{N}(0.273,0.030^2)$ & $0.1002 ^{+0.0041}_{-0.0017}$ & $0.1020 ^{+0.0009}_{-0.0009}$ & $0.1017 ^{+0.0009}_{-0.0011}$\\
    616--621 & $\mathcal{N}(0.397,0.050^2)$ & $\mathcal{N}(0.277,0.030^2)$ & $0.1020 ^{+0.0014}_{-0.0013}$ & $0.1014 ^{+0.0008}_{-0.0008}$ & $0.1016 ^{+0.0006}_{-0.0006}$\\
    622--626 & $\mathcal{N}(0.394,0.049^2)$ & $\mathcal{N}(0.276,0.028^2)$ & $0.1007 ^{+0.0017}_{-0.0017}$ & $0.1013 ^{+0.0018}_{-0.0016}$ & $0.1009 ^{+0.0012}_{-0.0011}$\\
    626--632 & $\mathcal{N}(0.389,0.050^2)$ & $\mathcal{N}(0.280,0.029^2)$ & $0.0952 ^{+0.0024}_{-0.0025}$ & $0.1030 ^{+0.0007}_{-0.0008}$ & $0.1025 ^{+0.0008}_{-0.0009}$\\
    632--636 & $\mathcal{N}(0.385,0.049^2)$ & $\mathcal{N}(0.279,0.029^2)$ & $0.1002 ^{+0.0020}_{-0.0020}$ & $0.0989 ^{+0.0016}_{-0.0013}$ & $0.0992 ^{+0.0014}_{-0.0011}$\\
    637--642 & $\mathcal{N}(0.380,0.049^2)$ & $\mathcal{N}(0.281,0.029^2)$ & $0.0977 ^{+0.0023}_{-0.0025}$ & $0.1002 ^{+0.0007}_{-0.0009}$ & $0.0999 ^{+0.0007}_{-0.0008}$\\
    642--646 & $\mathcal{N}(0.373,0.049^2)$ & $\mathcal{N}(0.283,0.029^2)$ & $0.1006 ^{+0.0018}_{-0.0019}$ & $0.1001 ^{+0.0006}_{-0.0006}$ & $0.1000 ^{+0.0006}_{-0.0006}$\\
    647--652 & $\mathcal{N}(0.366,0.049^2)$ & $\mathcal{N}(0.286,0.028^2)$ & $0.0979 ^{+0.0023}_{-0.0025}$ & $0.1002 ^{+0.0009}_{-0.0009}$ & $0.0999 ^{+0.0008}_{-0.0008}$\\
    652--656 & $\mathcal{N}(0.320,0.053^2)$ & $\mathcal{N}(0.315,0.029^2)$ & $0.0953 ^{+0.0028}_{-0.0028}$ & $0.1002 ^{+0.0013}_{-0.0019}$ & $0.0989 ^{+0.0017}_{-0.0022}$\\
    657--662 & $\mathcal{N}(0.311,0.051^2)$ & $\mathcal{N}(0.314,0.026^2)$ & $0.1003 ^{+0.0015}_{-0.0017}$ & $0.1005 ^{+0.0017}_{-0.0012}$ & $0.1003 ^{+0.0009}_{-0.0008}$\\
    662--666 & $\mathcal{N}(0.359,0.048^2)$ & $\mathcal{N}(0.286,0.028^2)$ & $0.1007 ^{+0.0040}_{-0.0037}$ & $0.1013 ^{+0.0007}_{-0.0008}$ & $0.1013 ^{+0.0007}_{-0.0008}$\\
    667--672 & $\mathcal{N}(0.358,0.048^2)$ & $\mathcal{N}(0.283,0.028^2)$ & $0.1000 ^{+0.0014}_{-0.0014}$ & $0.1011 ^{+0.0008}_{-0.0008}$ & $0.1008 ^{+0.0007}_{-0.0007}$\\
    672--676 & $\mathcal{N}(0.356,0.045^2)$ & $\mathcal{N}(0.282,0.026^2)$ & $0.0963 ^{+0.0022}_{-0.0026}$ & $0.1001 ^{+0.0013}_{-0.0012}$ & $0.0991 ^{+0.0010}_{-0.0011}$\\
    677--682 & $\mathcal{N}(0.354,0.045^2)$ & $\mathcal{N}(0.281,0.026^2)$ & $0.1009 ^{+0.0011}_{-0.0014}$ & $0.0996 ^{+0.0013}_{-0.0010}$ & $0.1002 ^{+0.0012}_{-0.0010}$\\
    682--686 & $\mathcal{N}(0.350,0.044^2)$ & $\mathcal{N}(0.281,0.025^2)$ & $0.0948 ^{+0.0041}_{-0.0037}$ & $0.1034 ^{+0.0010}_{-0.0010}$ & $0.1027 ^{+0.0009}_{-0.0016}$\\
    686--691 & $\mathcal{N}(0.347,0.044^2)$ & $\mathcal{N}(0.281,0.025^2)$ & $0.1004 ^{+0.0010}_{-0.0011}$ & $0.0989 ^{+0.0012}_{-0.0010}$ & $0.0999 ^{+0.0011}_{-0.0011}$\\
    692--696 & $\mathcal{N}(0.344,0.045^2)$ & $\mathcal{N}(0.281,0.025^2)$ & $0.1012 ^{+0.0014}_{-0.0027}$ & $0.0998 ^{+0.0013}_{-0.0011}$ & $0.1002 ^{+0.0013}_{-0.0012}$\\
    696--701 & $\mathcal{N}(0.342,0.045^2)$ & $\mathcal{N}(0.280,0.025^2)$ & $0.0970 ^{+0.0016}_{-0.0019}$ & $0.1008 ^{+0.0007}_{-0.0007}$ & $0.1001 ^{+0.0007}_{-0.0009}$\\
    702--706 & $\mathcal{N}(0.339,0.044^2)$ & $\mathcal{N}(0.280,0.025^2)$ & $0.1034 ^{+0.0010}_{-0.0011}$ & $0.0990 ^{+0.0016}_{-0.0016}$ & $0.1018 ^{+0.0012}_{-0.0017}$\\
    706--711 & $\mathcal{N}(0.334,0.045^2)$ & $\mathcal{N}(0.281,0.025^2)$ & $0.1020 ^{+0.0012}_{-0.0013}$ & $0.0996 ^{+0.0007}_{-0.0008}$ & $0.1002 ^{+0.0009}_{-0.0007}$\\
    712--716 & $\mathcal{N}(0.331,0.045^2)$ & $\mathcal{N}(0.281,0.025^2)$ & $0.1000 ^{+0.0015}_{-0.0016}$ & $0.1007 ^{+0.0011}_{-0.0014}$ & $0.1004 ^{+0.0009}_{-0.0011}$\\
    716--721 & $\mathcal{N}(0.329,0.043^2)$ & $\mathcal{N}(0.278,0.024^2)$ & $0.0970 ^{+0.0021}_{-0.0016}$ & $0.0988 ^{+0.0010}_{-0.0012}$ & $0.0984 ^{+0.0010}_{-0.0012}$\\
    722--726 & $\mathcal{N}(0.325,0.044^2)$ & $\mathcal{N}(0.281,0.024^2)$ & $0.0961 ^{+0.0022}_{-0.0024}$ & $0.0999 ^{+0.0013}_{-0.0013}$ & $0.0989 ^{+0.0011}_{-0.0011}$\\
    726--731 & $\mathcal{N}(0.323,0.043^2)$ & $\mathcal{N}(0.279,0.024^2)$ & $0.0978 ^{+0.0009}_{-0.0009}$ & $0.0975 ^{+0.0016}_{-0.0017}$ & $0.0979 ^{+0.0008}_{-0.0007}$\\
    732--736 & $\mathcal{N}(0.321,0.043^2)$ & $\mathcal{N}(0.281,0.024^2)$ & $0.1039 ^{+0.0013}_{-0.0014}$ & $0.1008 ^{+0.0023}_{-0.0024}$ & $0.1031 ^{+0.0012}_{-0.0013}$\\
    736--741 & $\mathcal{N}(0.317,0.040^2)$ & $\mathcal{N}(0.277,0.022^2)$ & $0.1014 ^{+0.0025}_{-0.0032}$ & $0.0994 ^{+0.0014}_{-0.0017}$ & $0.1000 ^{+0.0013}_{-0.0014}$\\
    742--746 & $\mathcal{N}(0.313,0.041^2)$ & $\mathcal{N}(0.279,0.022^2)$ & $0.1022 ^{+0.0011}_{-0.0013}$ & $0.0988 ^{+0.0012}_{-0.0010}$ & $0.1007 ^{+0.0015}_{-0.0015}$\\
    746--752 & $\mathcal{N}(0.310,0.042^2)$ & $\mathcal{N}(0.281,0.023^2)$ & $0.1002 ^{+0.0014}_{-0.0012}$ & $0.0992 ^{+0.0027}_{-0.0016}$ & $0.0995 ^{+0.0011}_{-0.0010}$\\
    752--756 & $\mathcal{N}(0.308,0.042^2)$ & $\mathcal{N}(0.282,0.023^2)$ & $0.0981 ^{+0.0021}_{-0.0020}$ & $0.0993 ^{+0.0011}_{-0.0011}$ & $0.0991 ^{+0.0009}_{-0.0009}$\\
    766--770 & $\mathcal{N}(0.298,0.041^2)$ & $\mathcal{N}(0.281,0.022^2)$ & $0.1016 ^{+0.0011}_{-0.0023}$ & $0.0986 ^{+0.0013}_{-0.0013}$ & $0.0995 ^{+0.0014}_{-0.0015}$\\
    771--776 & $\mathcal{N}(0.297,0.039^2)$ & $\mathcal{N}(0.280,0.021^2)$ & $0.1009 ^{+0.0023}_{-0.0033}$ & $0.0997 ^{+0.0021}_{-0.0017}$ & $0.1000 ^{+0.0020}_{-0.0015}$\\
    776--780 & $\mathcal{N}(0.293,0.039^2)$ & $\mathcal{N}(0.282,0.020^2)$ & $0.1045 ^{+0.0023}_{-0.0022}$ & $0.0986 ^{+0.0011}_{-0.0013}$ & $0.0998 ^{+0.0011}_{-0.0011}$\\
    781--786 & $\mathcal{N}(0.292,0.038^2)$ & $\mathcal{N}(0.280,0.020^2)$ & $0.1016 ^{+0.0012}_{-0.0013}$ & $0.0979 ^{+0.0013}_{-0.0015}$ & $0.0995 ^{+0.0012}_{-0.0012}$\\
    786--790 & $\mathcal{N}(0.291,0.041^2)$ & $\mathcal{N}(0.283,0.021^2)$ & $0.1021 ^{+0.0016}_{-0.0014}$ & $0.0993 ^{+0.0009}_{-0.0008}$ & $0.1003 ^{+0.0009}_{-0.0008}$\\
    791--796 & $\mathcal{N}(0.289,0.041^2)$ & $\mathcal{N}(0.281,0.022^2)$ & $0.1016 ^{+0.0028}_{-0.0023}$ & $0.1003 ^{+0.0034}_{-0.0017}$ & $0.1006 ^{+0.0022}_{-0.0014}$\\
    796--800 & $\mathcal{N}(0.288,0.041^2)$ & $\mathcal{N}(0.282,0.022^2)$ & $0.1064 ^{+0.0038}_{-0.0042}$ & $0.0984 ^{+0.0008}_{-0.0010}$ & $0.0989 ^{+0.0008}_{-0.0008}$\\
    801--806 & $\mathcal{N}(0.285,0.041^2)$ & $\mathcal{N}(0.281,0.022^2)$ & $0.0995 ^{+0.0018}_{-0.0023}$ & $0.1001 ^{+0.0011}_{-0.0012}$ & $0.0999 ^{+0.0010}_{-0.0010}$\\
    806--810 & $\mathcal{N}(0.285,0.041^2)$ & $\mathcal{N}(0.280,0.022^2)$ & $0.1003 ^{+0.0017}_{-0.0022}$ & $0.0976 ^{+0.0023}_{-0.0029}$ & $0.0994 ^{+0.0013}_{-0.0016}$\\
    810--815 & $\mathcal{N}(0.283,0.043^2)$ & $\mathcal{N}(0.282,0.024^2)$ & $0.0989 ^{+0.0013}_{-0.0014}$ & $0.0986 ^{+0.0013}_{-0.0017}$ & $0.0989 ^{+0.0009}_{-0.0010}$\\
    816--820 & $\mathcal{N}(0.278,0.041^2)$ & $\mathcal{N}(0.280,0.022^2)$ & $0.0969 ^{+0.0024}_{-0.0021}$ & $0.0985 ^{+0.0013}_{-0.0010}$ & $0.0983 ^{+0.0009}_{-0.0009}$\\
    820--825 & $\mathcal{N}(0.275,0.039^2)$ & $\mathcal{N}(0.279,0.021^2)$ & $0.0950 ^{+0.0009}_{-0.0009}$ & $0.1022 ^{+0.0009}_{-0.0009}$ & $0.1007 ^{+0.0013}_{-0.0048}$\\
    826--830 & $\mathcal{N}(0.275,0.039^2)$ & $\mathcal{N}(0.281,0.021^2)$ & $0.0984 ^{+0.0024}_{-0.0019}$ & $0.0990 ^{+0.0008}_{-0.0007}$ & $0.0991 ^{+0.0007}_{-0.0007}$\\
    830--835 & $\mathcal{N}(0.274,0.040^2)$ & $\mathcal{N}(0.278,0.021^2)$ & $0.0982 ^{+0.0010}_{-0.0011}$ & $0.0970 ^{+0.0019}_{-0.0021}$ & $0.0980 ^{+0.0009}_{-0.0010}$\\
    836--840 & $\mathcal{N}(0.272,0.039^2)$ & $\mathcal{N}(0.278,0.021^2)$ & $0.0965 ^{+0.0012}_{-0.0012}$ & $0.0993 ^{+0.0030}_{-0.0028}$ & $0.0969 ^{+0.0011}_{-0.0010}$\\
    840--851 & $\mathcal{N}(0.264,0.039^2)$ & $\mathcal{N}(0.281,0.020^2)$ & $0.0994 ^{+0.0024}_{-0.0020}$ & $0.0976 ^{+0.0009}_{-0.0009}$ & $0.0979 ^{+0.0008}_{-0.0007}$\\
    852--862 & $\mathcal{N}(0.258,0.039^2)$ & $\mathcal{N}(0.281,0.020^2)$ & $0.1001 ^{+0.0007}_{-0.0007}$ & $0.0998 ^{+0.0019}_{-0.0018}$ & $0.1001 ^{+0.0006}_{-0.0006}$\\
    862--874 & $\mathcal{N}(0.247,0.032^2)$ & $\mathcal{N}(0.271,0.015^2)$ & $0.0990 ^{+0.0019}_{-0.0019}$ & $0.0988 ^{+0.0012}_{-0.0015}$ & $0.0990 ^{+0.0010}_{-0.0011}$\\
    874--884 & $\mathcal{N}(0.253,0.037^2)$ & $\mathcal{N}(0.280,0.019^2)$ & $0.0975 ^{+0.0020}_{-0.0015}$ & $0.0973 ^{+0.0010}_{-0.0011}$ & $0.0974 ^{+0.0008}_{-0.0009}$\\
    885--896 & $\mathcal{N}(0.256,0.037^2)$ & $\mathcal{N}(0.279,0.019^2)$ & $0.0972 ^{+0.0010}_{-0.0010}$ & $0.0974 ^{+0.0017}_{-0.0018}$ & $0.0972 ^{+0.0008}_{-0.0008}$\\
    896--906 & $\mathcal{N}(0.245,0.033^2)$ & $\mathcal{N}(0.271,0.016^2)$ & $0.0982 ^{+0.0035}_{-0.0032}$ & $0.1004 ^{+0.0011}_{-0.0012}$ & $0.1001 ^{+0.0009}_{-0.0011}$\\
    907--938 & $\mathcal{N}(0.248,0.037^2)$ & $\mathcal{N}(0.281,0.019^2)$ & $0.0977 ^{+0.0028}_{-0.0030}$ & $0.0977 ^{+0.0020}_{-0.0024}$ & $0.0980 ^{+0.0017}_{-0.0019}$\\
\enddata
\end{deluxetable}

\begin{deluxetable}{lcccc}
\tablecaption{Parameter estimation for spectral retrievals. \label{tab:retrievalparam}}
\tablewidth{0pt}
\tabletypesize{\footnotesize}
\tablehead{
\colhead{Parameter} & 
\colhead{Prior} & 
\multicolumn{3}{c}{Posterior estimate} \\
\colhead{} & 
\colhead{} & 
\colhead{2018-07-29} &
\colhead{2020-08-07} &
\colhead{Joint}
} 
\startdata
    \multicolumn{5}{l}{\it\ \ Retrieval assuming equilibrium chemistry}\\
    T-P $\log\delta$(bar$^{-1}$)            & $\mathcal{N}(-5.5,2.5^2)$      & $-6.6^{+1.3}_{-1.8}$ & $-6.4^{+1.8}_{-1.7}$ & $-6.1^{+1.4}_{-1.6}$\\
    T-P $\log\gamma$                        & $\mathcal{N}(0,2^2)$           & $-1.4^{+0.9}_{-1.2}$ & $-0.9^{+1.2}_{-1.4}$ & $-1.0^{+1.1}_{-1.2}$\\
    T-P $T_\mathrm{int}$(K)                 & $\mathcal{U}(0,1500)$          & $974^{+351}_{-467}$ & $895^{+378}_{-477}$ & $789^{+415}_{-430}$\\
    T-P $T_\mathrm{eq}$(K)                  & $\mathcal{U}(0,4000)$          & $1612^{+84}_{-191}$ & $1774^{+339}_{-370}$ & $1645^{+255}_{-244}$\\
    T-P $\log P_\mathrm{trans}$(bar)        & $\mathcal{N}(-3,3^2)$          & $-5.1^{+1.4}_{-1.0}$ & $-3.9^{+2.4}_{-2.2}$ & $-3.8^{+1.9}_{-1.9}$\\
    T-P $\alpha$                            & $\mathcal{N}(0.25,0.4^2)$      & $0.35^{+0.28}_{-0.21}$ & $0.26^{+0.28}_{-0.28}$ & $0.26^{+0.27}_{-0.29}$\\
    $\log P_\mathrm{0}$(bar)                & $\mathcal{U}(-4,2)$            & $-3.7^{+0.3}_{-0.2}$ & $-3.6^{+0.4}_{-0.2}$ & $-3.5^{+0.5}_{-0.3}$\\
    $\log P_\mathrm{cloud}$(bar)            & $\mathcal{U}(-4,2)$            & $-0.5^{+1.5}_{-1.7}$ & $-1.2^{+1.7}_{-1.4}$ & $-1.6^{+1.8}_{-1.5}$\\
    $\log A_\mathrm{RS}$                    & $\mathcal{U}(0,4)$             & $1.2^{+0.8}_{-0.7}$ & $1.6^{+1.0}_{-0.9}$ & $1.7^{+0.9}_{-0.8}$\\
    C/O                                     & $\mathcal{U}(0.05,2)$          & $0.7^{+0.1}_{-0.2}$ & $1.4^{+0.3}_{-0.3}$ & $1.2^{+0.4}_{-0.6}$\\
    $\log Z$                                & $\mathcal{U}(-2,3)$            & $0.4^{+0.6}_{-0.9}$ & $0.9^{+0.6}_{-1.3}$ & $0.0^{+1.1}_{-1.4}$\\
    $\phi$                                  & $\mathcal{U}(0,1)$             & $0.37^{+0.32}_{-0.25}$ & $0.37^{+0.24}_{-0.19}$ & $0.36^{+0.23}_{-0.17}$\\
    \hline\noalign{\smallskip}
    \multicolumn{5}{l}{\it\ \ Retrieval assuming free chemistry}\\
    T-P $T_\mathrm{iso}$(K)                 & $\mathcal{U}(800,3000)$        & $1170^{+266}_{-233}$ & $2410^{+402}_{-614}$ & $1018^{+225}_{-138}$\\
    $\log P_\mathrm{0}$(bar)                & $\mathcal{U}(-4,2)$            & $-3.3^{+0.7}_{-0.4}$ & $-3.6^{+0.4}_{-0.3}$ & $-3.4^{+0.6}_{-0.4}$\\
    $\log P_\mathrm{cloud}$(bar)            & $\mathcal{U}(-4,2)$            & $0.5^{+1.0}_{-1.1}$ & $-2.7^{+0.5}_{-0.3}$ & $0.2^{+1.2}_{-1.3}$\\
    $\log A_\mathrm{RS}$                    & $\mathcal{U}(-2,4)$            & $-0.4^{+1.0}_{-1.0}$ & $-0.4^{+1.0}_{-1.0}$ & $0.3^{+0.7}_{-1.0}$\\
    $\log X_\mathrm{TiO}$                   & $\mathcal{U}(-10,0)$           & $-8.5^{+0.6}_{-0.8}$ & $-7.4^{+0.3}_{-0.5}$ & $-8.2^{+0.5}_{-0.6}$\\
    $\log X_\mathrm{VO}$                    & $\mathcal{U}(-10,0)$           & $-7.9^{+0.6}_{-0.7}$ & $-9.1^{+0.7}_{-0.6}$ & $-8.5^{+0.5}_{-0.7}$\\
    $\log X_\mathrm{Na}$                    & $\mathcal{U}(-10,0)$           & $-8.0^{+1.5}_{-1.3}$ & $-3.3^{+0.7}_{-1.1}$ & $-5.6^{+1.1}_{-1.7}$\\
    $\log X_\mathrm{K}$                     & $\mathcal{U}(-10,0)$           & $-6.5^{+1.6}_{-1.5}$ & $-8.2^{+1.4}_{-1.2}$ & $-8.1^{+1.4}_{-1.2}$\\
    $\log X_\mathrm{H_2O}$                  & $\mathcal{U}(-10,0)$           & $-6.8^{+2.1}_{-2.2}$ & $-6.0^{+2.7}_{-2.7}$ & $-6.4^{+2.5}_{-2.3}$\\
    \hline\noalign{\smallskip}
    \multicolumn{5}{l}{\it\ \ Retrieval assuming free chemistry with spots/faculae contamination}\\
    T-P $T_\mathrm{iso}$(K)                 & $\mathcal{U}(800,3000)$        & $1241^{+245}_{-239}$ & $1900^{+545}_{-341}$ & $1091^{+224}_{-166}$\\
    $\log P_\mathrm{0}$(bar)                & $\mathcal{U}(-4,2)$            & $-3.5^{+0.6}_{-0.3}$ & $-3.1^{+0.8}_{-0.5}$ & $-3.0^{+0.8}_{-0.6}$\\
    $\log P_\mathrm{cloud}$(bar)            & $\mathcal{U}(-4,2)$            & $0.5^{+1.0}_{-1.0}$ & $-0.6^{+1.3}_{-1.2}$ & $0.0^{+1.1}_{-1.2}$\\
    $\log A_\mathrm{RS}$                    & $\mathcal{U}(-2,4)$            & $-0.6^{+0.9}_{-0.8}$ & $0.2^{+1.2}_{-1.1}$ & $0.2^{+0.9}_{-1.1}$\\
    $\log X_\mathrm{TiO}$                   & $\mathcal{U}(-10,0)$           & $-8.8^{+0.8}_{-0.6}$ & $-7.7^{+0.7}_{-0.9}$ & $-7.7^{+0.6}_{-0.7}$\\
    $\log X_\mathrm{VO}$                    & $\mathcal{U}(-10,0)$           & $-8.2^{+0.7}_{-0.5}$ & $-9.0^{+0.7}_{-0.6}$ & $-8.4^{+0.7}_{-0.7}$\\
    $\log X_\mathrm{Na}$                    & $\mathcal{U}(-10,0)$           & $-8.2^{+1.6}_{-1.2}$ & $-1.0^{+0.5}_{-0.9}$ & $-2.5^{+1.3}_{-1.5}$\\
    $\log X_\mathrm{K}$                     & $\mathcal{U}(-10,0)$           & $-6.8^{+1.4}_{-1.1}$ & $-8.0^{+1.4}_{-1.2}$ & $-7.7^{+1.6}_{-1.3}$\\
    $\log X_\mathrm{H_2O}$                  & $\mathcal{U}(-10,0)$           & $-6.7^{+1.8}_{-2.0}$ & $-5.8^{+2.4}_{-2.3}$ & $-5.9^{+2.3}_{-2.4}$\\
    $T_\mathrm{spot}$(K)                    & $\mathcal{U}(2000,5835)$       & $5358^{+290}_{-639}$ & $4292^{+867}_{-656}$ & $4930^{+508}_{-888}$\\
    $f_\mathrm{spot}$                       & $\mathcal{U}(0,1)$             & $0.15^{+0.23}_{-0.10}$ & $0.13^{+0.10}_{-0.07}$ & $0.10^{+0.15}_{-0.06}$\\
    $T_\mathrm{faculae}$(K)                 & $\mathcal{U}(5835,7000)$       & $5997^{+214}_{-107}$ & $6257^{+251}_{-128}$ & $6116^{+134}_{-102}$\\
    $f_\mathrm{faculae}$                    & $\mathcal{U}(0,1)$             & $0.25^{+0.35}_{-0.17}$ & $0.50^{+0.24}_{-0.20}$ & $0.53^{+0.23}_{-0.19}$\\
\enddata
\end{deluxetable}

\begin{deluxetable}{ccc|cccc|cccc|cccc}
\tablecaption{Statistics from Bayesian spectral retrieval analysis. \label{tab:retrievalstat}}
\tablewidth{0pt}
\tabletypesize{\footnotesize}
\tablehead{
\colhead{\#} & 
\colhead{Model} & 
\colhead{dof} & 
\multicolumn{4}{c}{2018-07-29} & 
\multicolumn{4}{c}{2020-08-07} & 
\multicolumn{4}{c}{Joint}  \\
\colhead{} & 
\colhead{} & 
\colhead{} & 
\colhead{$\chi^2_\mathrm{MAP}$\tablenotemark{a}} & 
\colhead{$\ln\mathcal{Z}$} & 
\colhead{$\Delta\ln\mathcal{Z}$} & 
\colhead{FS\tablenotemark{b}} & 
\colhead{$\chi^2_\mathrm{MAP}$\tablenotemark{a}} & 
\colhead{$\ln\mathcal{Z}$} & 
\colhead{$\Delta\ln\mathcal{Z}$} & 
\colhead{FS\tablenotemark{b}} & 
\colhead{$\chi^2_\mathrm{MAP}$\tablenotemark{a}} & 
\colhead{$\ln\mathcal{Z}$} & 
\colhead{$\Delta\ln\mathcal{Z}$} & 
\colhead{FS\tablenotemark{b}}
} 
\startdata
\hline\noalign{\smallskip}
    \multicolumn{15}{l}{\it\ \ A. Simple assumption}\\
    1 & Flat line       & 69 & 135.4 & 410.98 & $-$0.3  & N/A
                        & 132.1 & 442.09 & $-$6.9  & 4.1$\sigma$
                        & 127.7 & 456.23 & $-$10.0 & 4.8$\sigma$\\
    2 & Sloped line     & 68 & 127.8 & 411.25 & 0       & Ref.
                        & 111.1 & 448.94 & 0       & Ref.
                        & 99.1 & 466.20 & 0       & Ref.\\
    \hline\noalign{\smallskip}
    \multicolumn{15}{l}{\it\ \ B. Retrieval assuming equilibrium chemistry}\\
    1 & Full model      & 58 & 107.1 & 419.05 & 0       & Ref.
                        & 93.2 & 457.82 & 0       & Ref.
                        & 77.9 & 475.79 & 0       & Ref.\\
    \hline\noalign{\smallskip}
    \multicolumn{15}{l}{\it\ \ C. Retrieval assuming free chemistry}\\
    1 & Full model      & 61 & 103.1 & 423.22 & 0      & Ref.
                        & 83.8 & 459.48 & 0      & Ref.
                        & 76.6 & 476.34 & 0      & Ref.\\
    2 & No TiO          & 60 & 109.9 & 421.12 & $-$2.1 & 2.6$\sigma$
                        & 94.9 & 455.62 & $-$3.9 & 3.3$\sigma$
                        & 89.2 & 471.14 & $-$5.2 & 3.7$\sigma$\\
    3 & No VO           & 60 & 115.1 & 418.49 & $-$4.7 & 3.5$\sigma$
                        & 84.0 & 460.78 & 1.3 & $-$2.2$\sigma$
                        & 81.9 & 475.41 & $-$0.9 & 2.0$\sigma$\\
    4 & No Na           & 60 & 102.4 & 423.96 & 0.7 & N/A
                        & 92.3 & 456.26 & $-$3.2 & 3.0$\sigma$
                        & 79.5 & 475.18 & $-$1.2 & 2.1$\sigma$\\
    5 & No K            & 60 & 105.2 & 422.03 & $-$1.2 & 2.1$\sigma$
                        & 83.8 & 460.05 & 0.6    & N/A
                        & 76.8 & 477.05 & 0.7    & N/A\\
    6 & No H$_2$O       & 60 & 102.4 & 423.60 & 0.4 & N/A
                        & 84.1 & 458.82 & $-$0.7 & N/A
                        & 76.4 & 476.04 & $-$0.3 & N/A\\
    7 & No TiO+Na       & 59 & 110.1 & 421.68 & $-$1.5 & 2.3$\sigma$
                        & 112.7 & 449.50 & $-$10.0 & 4.8$\sigma$
                        & 98.8 & 468.08 & $-$8.3 & 4.5$\sigma$\\
    \hline\noalign{\smallskip}
    \multicolumn{15}{l}{\it\ \ D. Retrieval assuming spots/faculae contamination}\\
    1 & C1 with spots/faculae & 57 & 99.6 & 420.70 & 0      & Ref.
                              & 72.7 & 463.09 & 0      & Ref.
                              & 73.9 & 474.98 & 0      & Ref.\\
    2 & Spots/faculae only    & 65 & 106.7 & 417.84 & $-$2.9 & 2.9$\sigma$ 
                              & 93.4 & 457.08 & $-$6.0 & 3.9$\sigma$ 
                              & 86.4 & 471.97 & $-$3.0 & 3.0$\sigma$ \\
\enddata
\tablenotetext{a}{$\chi^2$ for the maximum a posteriori (MAP) model.}
\tablenotetext{b}{The Bayes factor $B_{10}=\mathcal{Z}_1/\mathcal{Z}_0$, or $\ln B_{10}=\Delta\ln\mathcal{Z}$, was converted to frequentist significance (FS) following \citet{2008ConPh..49...71T} and \citet{2013ApJ...778..153B}. It is labeled as N/A if $|\Delta\ln\mathcal{Z}|<0.9$. The number of ``sigmas'' is a useful frequentist metric to express the odds in favor of a more complex model, which does not necessarily mean the detection of a species in the spectrum of a planet \citep{2021arXiv210308600W}.
}
\end{deluxetable}

\bibliography{ref_db.bib}{}
\bibliographystyle{aasjournal}



\end{document}